\documentclass[journal,lettersize]{IEEEtran}
\usepackage{cite}
\usepackage{graphicx}
\usepackage{colortbl}
\usepackage{amsmath,amssymb,bm}
\usepackage{stfloats}
\usepackage{array}
\usepackage{tabularx}
\usepackage{booktabs}
\usepackage[table,xcdraw]{xcolor}
\usepackage{hhline}
\ifCLASSOPTIONcompsoc
  \usepackage[caption=false,font=normalsize,labelfont=sf,textfont=sf]{subfig}
\else
  \usepackage[caption=false,font=footnotesize]{subfig}
\fi
\usepackage{multirow}
\usepackage{xurl}
\usepackage{bbm}
\usepackage{nicematrix}
\usepackage{balance}

\IEEEoverridecommandlockouts
\begin{document}
\title{SBMA: A Multiple Access Scheme Combining SCMA and BIA for MU-MISO}
\author{Jianjian~Wu, Chi-Tsun Cheng,~\IEEEmembership{Senior Member,~IEEE}, Qingfeng~Zhou,~\IEEEmembership{Member,~IEEE}, Jianlin Liang, Jinke Wu
	\thanks{This work was supported in part by NSFC under Grant 61971138, and in part by Guangdong Basic and Applied Basic Research Foundation 2022A1515140137. (Corresponding author: Qingfeng~Zhou)
	\par J. Wu and Q.F. Zhou are with the School of Electric Engineering and Intelligentization, Dongguan University of Technology, Dongguan 523808, China, {and with Guangdong Provincial Key Laboratory of Intelligent Disaster Prevention and Emergency Technologies for Urban Lifeline Engineering}, and with the Dongguan Key Lab of Artificial Information Network.  C.-T. Cheng is with the Department of Manufacturing, Materials and Mechatronics, RMIT University, Melbourne, VIC 3000, Australia (e-mail:ben.cheng@rmit.edu.au). J.L. Liang and J.K. Wu are with China Mobile Group Guangdong Co, Ltd. Dongguan Branch.
		}
}
\maketitle
\begin{abstract}
Sparse Code Multiple Access (SCMA) and Blind Interference Alignment (BIA) are key enablers for multi-user communication, yet each suffers from distinct limitations: SCMA faces high complexity and limited multiplexing gain, while BIA requires a long temporal channel pattern and incurs significant decoding delay. This paper proposes SBMA (Sparsecode-and-BIA-based Multiple Access), a novel framework that synergizes SCMA’s diversity and BIA’s multiplexing while addressing their drawbacks. We design two decoders: a low-complexity two-stage decoder (Zero-forcing + Message Passing Algorithm (MPA)) and a Joint MPA (JMPA) decoder leveraging a virtual factor graph for improved BER. Theoretical analysis derives closed-form BER expressions for a 6-user $2\times 1$ MISO system, validated by simulations. Compared to existing schemes, SBMA with JMPA achieves a diversity gain equivalent to STBC-SCMA and a multiplexing gain comparable to BIA, while simultaneously offering enhanced privacy (relative to STBC-SCMA) and reduced reliance on channel coherence time (compared to BIA). These advancements position SBMA as a compelling solution for next-generation wireless communication systems, particularly in IoT applications demanding high throughput, robust data privacy, and adaptability to dynamic channel conditions.
\end{abstract}



\section{Introduction}
\label{Sect.Intro}
The Internet of Things (IoT) is rapidly evolving to bridge physical environments with cyber systems, enabling transformative applications such as autonomous factories, smart cities, and connected healthcare. By the end of 2023, there were 16.6 billion connected IoT devices, with an estimated increase to 40 billion by 2030 \cite{IoT2024}. The rapid growth of IoT devices, expected to exceed one million per square kilometer, will generate unprecedented data volumes and connectivity demands, surpassing the capabilities of 5G technology, particularly for applications requiring ultra-low latency and terabit-level data rates \cite{SecMassiveIoT2023}.

To address these challenges, Interference Alignment (IA)\cite{2008Cadambe_IAandDoF} and Non-Orthogonal Multiple Access (NOMA)\cite{2017Ding_AppOfNoma,2021Yuan_NOMAforNGMIoT} have emerged as key enablers.

\subsection{Related works and Motivation}
IA aligns multiple interference signal vectors into a small subspace and reserves the remaining subspace for transmitting the intended signal\cite{2008Maddah_ComOverMIMOX_IA}. {Given its remarkable capacity to manage a large number of interference channels simultaneously, IA has been applied to various communication scenarios \cite{2016NZhao_IASurvey}, including cognitive radio networks\cite{2015NZhao_AdapPAInIACognitiveNet}, cellular networks\cite{2008SuH_IAforCellularNet}, heterogeneous networks\cite{2012Shin_HIAforDownlinkHetNet}, and Device-to-Device (D2D) networks\cite{2012Elkotby_ExploitingIAInD2DCellNet}.} A primary limitation throughout the development of IA is the requirement for channel state information at the transmitter (CSIT)\cite{2014Razavi_PerformAnalysisOfIAunderCSImismatch}. To alleviate the dependency on CSIT, blind IA (BIA) was introduced for multiple access channels \cite{Gou2011AimingPerfectlyintheDarkBIA}. BIA operates via predefined channel patterns and is implemented through two methods: channel-block-based BIA (CB-BIA)\cite{Zhou2013DiophantineApproachtoBIA} and reconfigurable-antenna-based BIA (RA-BIA). {BIA performs transmissions using specially designed channel patterns over an extension of several symbol extensions (i.e., timeslots), often referred to as a ``supersymbol''.} CB-BIA schemes leverage distinctive features of homogeneous block fading channels to fulfill supersymbols\cite{Zhou2013DiophantineApproachtoBIA}. The RA-BIA accomplishes supersymbols by employing reconfigurable antennas (RA), enabling the artificial modification of the channel state. The mode-switching cost related to RA-BIA has been well studied in \cite{zhou2017OntheModeSwitchingofBIA} and \cite{JJWu2020BalancedSwitchingBIA}. A comprehensive survey on BIA is provided in \cite{U2022BIASurvey}. Despite BIA providing high degrees of freedom in the absence of CSIT, it necessitates the channel coefficients of all users to remain constant over the entire signal processing block. However, the length of supersymbols in RA-BIA is approximately proportional to $M^K$, where $M$ is the number of transmit antennas (or preset modes) and $K$ is the number of users. This implies that the required length of channel coherence time may not be feasible in practice. Scholars have proposed some BIA designs under limited symbol extensions, but these designs impose constraints that must be satisfied \cite{Yang2016HierarchicalBIA} and become increasingly complicated as $M$ and $K$ grow \cite{Cha2018BIAfortheKuserMISOBC}.

NOMA utilizes superposition coding to allow multicasting signals to multiple users on the same resource block (e.g., time and frequency). There are two major types of NOMA, namely Power-Domain NOMA (PD-NOMA) and Code-Domain NOMA (CD-NOMA)\cite{2015Dai_NOMAfor5G}. 
Sparse code multiple access (SCMA) is a well-regarded CD-NOMA scheme that has attracted significant attention \cite{Nikopour2013SCMA}.
In multi-carrier systems, SCMA employs low-density multi-dimensional codebooks and message passing algorithms (MPA), which offer high spectrum efficiency and unique ``shaping gain'' \cite{Taherzadeh2014SCMACodebookDesign}. However, the diversity order of SCMA systems is limited by the sparsity assigned to users, i.e., the effective number of resource nodes occupied by each user \cite{Liu2021SparseorDense}. To improve the diversity gain of downlink SCMA systems, Luo {\em et al.} introduced the quadrature component delay into the codeword, achieving additional signal space diversity \cite{Luo2023EnhancingSSDforSCMA}. Another straightforward way to improve diversity is to integrate MIMO into SCMA systems. In MIMO-SCMA systems, joint MPA decoding is commonly utilized to achieve performance that is close to maximum likelihood decoding while maintaining low complexity \cite{Tang2016JMPA}. Lim {\em et al.} studied an uplink SCMA system with multiple receive antennas, proposing a Gaussian approximation-based BP (GA-BP) detection with linear prefiltering to achieve space diversity gain \cite{Lim2017UplinkSCMAwithMA}. Besides, Pan et al. incorporated space-time block coding (STBC) into MIMO-SCMA systems, known as STBC-based SCMA that achieves transmit diversity gain \cite{Pan2019STBC}. Chen {\em et al.} investigated the designs of optimal codebook and receiver detector for MIMO-SCMA systems, achieving spatial and frequency resource diversity \cite{Chen2022MIMOSCMA}. While these studies primarily focus on achieving higher diversity order for MIMO-SCMA systems, less work has been done on enhancing the multiplexing gain. The asymptotic sum rate of a MIMO-SCMA has been calculated from asymptotic SINR using random matrix theory, and proven to achieve higher multiplexing performance than MIMO-OFDMA in \cite{LiuT2015CapacityForSCMA}, but without providing a feasible transceiver design. To the best of our knowledge, none of the previous downlink MIMO-SCMA schemes achieve spatial multiplexing gain with affordable decoding complexity.
Besides the above-mentioned concerns, another limitation of existing SCMA schemes is the opening for privacy breaches in multi-user systems. In such systems, MPA enables receivers to decode both desired and interference signals, potentially allowing unauthorized access to other users' data.

While PD-NOMA improves spectral efficiency via superposition coding, it suffers from error propagation in fading channels. BIA eliminates multi-user interference without CSIT but requires prohibitively long supersymbols. Conversely, SCMA achieves "shaping gain" via low-density codebooks but faces limited multiplexing capability and privacy risks due to shared codebook decoding. Despite recent advances in combining BIA with PD-NOMA\cite{Maximo2020BIANOMA}, which does not consider the maximization of multiplexing and diversity as does in this paper, no prior work has explored the integration of BIA and CD-NOMA (e.g., SCMA). This gap critically limits the potential to simultaneously achieve: 
(1) High multiplexing gain (via BIA’s interference-free spatial streams),  
(2) Enhanced diversity (via SCMA’s multi-tone coding),  
and (3) Practical deployment (reduced CSIT dependency and supersymbol length), thus motivating our work.

\subsection{Contributions}

This paper proposes SBMA (Sparsecode-and-BIA-based Multiple Access) — a novel multiple access scheme that synergistically combines BIA and SCMA through three key innovations:  

\begin{enumerate}[]
\item \textbf{Shared Codebook Architecture:} Eliminates user-specific codebooks by leveraging BIA’s interference cancellation, reducing codebook design complexity compared to conventional SCMA.

\item \textbf{Hierarchical Encoding:} The proposed hybrid framework optimally combines BIA's spatial multiplexing and SCMA's diversity gain, with adaptive stream allocation enabling flexible trade-offs between data streams and coherence time requirements.

\item \textbf{Privacy-Aware Decoding:} The integrated BIA encoding with corresponding interference cancellation prevents unauthorized data access, mitigating information leakage compared to standalone SCMA. 
\end{enumerate} 

Comprehensive evaluations in a 6-user $2\times 1$ MISO system with 4 subcarriers demonstrate that SBMA achieves:  
\begin{itemize}[]
\item \textbf{Diversity order of 4}, equivalent to STBC-SCMA,  
\item \textbf{Multiplexing gain of $\frac{48}{7}$}, matching BIA's theoretical limit,  
\item \textbf{Data streams of $\frac{72}{7}$}, surpassing BIA ($\frac{48}{7}$) and STBC-SCMA ($6$),
\item \textbf{Channel coherence time of 7 slots}, compatible with conventional BIA, 
\item \textbf{Reduced coherence time requirements} versus BIA for equivalent stream requirement,
\item \textbf{Zero inter-user data leakage} through guaranteed subspace orthogonality among users.
\end{itemize}

These innovations position SBMA as a scalable and secure multiple access solution tailored for IoT networks requiring high spectral efficiency under relaxed latency constraints.

\subsection{Notations and Organization}

For clarity, key notations and their definitions are summarized in Table~\ref{tab:notations}.

The remainder of this paper is structured as follows: Section~II presents the system model, followed by a review of BIA and STBC-SCMA in Section~III. Sections~IV and V elaborate on SBMA’s encoding/decoding framework and bit error rate (BER) analysis, respectively. Performance evaluations and comparisons are provided in Sections~VI and VII, with conclusions drawn in Section~VIII.

\begin{table}[!ht]
	\centering
	\caption{Notations and Definitions}
	\label{tab:notations}
	\begin{tabularx}{\columnwidth}{lX} 
		\toprule
		\textbf{Symbol} & \textbf{Definition} \\
		\midrule
		$\left| \mathbf{a} \right|$ & The Euclidean norm of vector $\mathbf{a}$. \\
		$\left\Vert \mathbf{A} \right\Vert_F$ & The Frobenius norm of matrix $\mathbf{A}$. \\
		$\left\Vert \mathcal{A} \right\Vert$ & The cardinality of set $\mathcal{A}$. \\
		$\left(\cdot\right)^*$ & The conjugate operation. \\
		$\left(\cdot\right)^T$ & The transpose operation. \\
		$\left(\cdot\right)^H$ & The Hermitian operation. \\
		$\otimes$ & The Kronecker product. \\
		$diag(\mathbf{a})$ & A diagonal matrix with vector $\mathbf{a}$ as its diagonal elements. \\
		$\mathbf{B} = blkdiag\{\mathbf{A}_1,\cdots, \mathbf{A}_N\}$ & A block diagonal matrix formed by aligning matrices $\mathbf{A}_1,\cdots, \mathbf{A}_N$ along the diagonal of $\mathbf{B}$. \\
		$\mathbb{E}(\cdot)$ & The expected value. \\
		$\mathcal{O}(\cdot)$ & The asymptotic upper bound describing the limiting behavior of a function as its argument tends toward infinity. \\
		\bottomrule
	\end{tabularx}
\end{table}

\section{System model}
\label{sysmodel}
\begin{figure}[ht]
	\centering
	\includegraphics[width=3.0in]{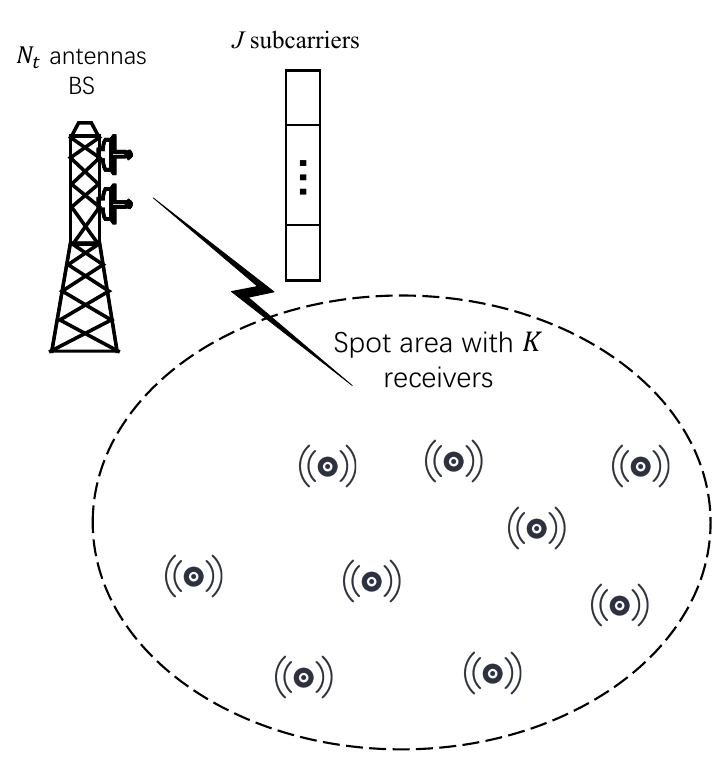}
	\caption{System model.}
	\label{Sys_model}
\end{figure}

To address the massive connectivity requirements of IoT networks, we study a $K$-user $N_t \times 1$ MISO channel with $J$ tones, as illustrated in Fig.~\ref{Sys_model}. The base station (BS) employs $N_t$ conventional antennas, while each receiver/user incorporates a reconfigurable antenna (RA).

The $K$ RAs are assumed to possess the capability to switch  between $N_t$ operational modes, thereby facilitating the artificial modification of the channel state \cite{Gou2011AimingPerfectlyintheDarkBIA,U2022BIASurvey}.

Specifically, we focus on a scenario with $K = 6$ users and $J = 4$ tones, which serves as a proof-of-concept for the proposed SBMA scheme in IoT applications with moderate user density and subcarrier availability. The composite channel model accounts for both quasi-static Rayleigh fading and large-scale path loss, reflecting practical wireless conditions. We denote the channel matrix from the BS to user $k $ on tone $j$ with RA mode $m=\mathbbmtt{1},\mathbbmtt{2},\dots$ $\mathbbmtt{N_t}$ as $\mathbf{h}_k^j(m) = \frac{\mathbf{g}_k^j(m)}{D(d_k)}$, where $\mathbf{g}_k^j(m) \in \mathbb{C}^{1 \times N_t}$ is the small-scale Rayleigh fading vector and $D(d_k)$ is the path loss function defined as
\begin{equation*}
D(d_k)=\left\{
\begin{aligned}
{d_k}^{\alpha}&,& & {\rm{if}} \quad d_k>r_0 & \\
{r_0}^{\alpha}&,& & \rm{otherwise} &
\end{aligned}
\right.
,
\nonumber
\end{equation*}
where $\alpha$ denotes the path loss exponent and the parameter $r_0$ avoids a singularity when the distance is small. Without loss of generality, we choose $\alpha=3$, which is used for the urban environment.
We assume no CSIT, full CSI at the receivers, and a sufficiently long coherence time to support the transmission process.
Note here $\mathbbmtt{N_t}=N_t$. The distinct font styles are employed to facilitate a more straightforward comprehension of the channel mode when it is presented.

\section{Related schemes}
\label{Sect.FormerSche}
Considering the aforementioned system model, we review two existing schemes in this section: BIA and STBC-based SCMA.

\subsection{BIA}
\label{modelBIA}

\begin{table}[ht]
	\centering
	\renewcommand\arraystretch{2.2} 
	\setlength{\tabcolsep}{4pt}
	\caption{Supersymbol of BIA}
	\label{fig:biasupersymbol}
		\begin{tabular}{cccccccc}
		\rowcolor[HTML]{FFFFFF} 
		& slot-1                                                             & slot-2                                                             & slot-3                                                             & slot-4                                                             & slot-5                                                             & slot-6                                                             & slot-7                                                             \\ \hhline{~|-------|}
		\rowcolor[HTML]{FFFFFF} 
		\multicolumn{1}{l|}{\cellcolor[HTML]{FFFFFF}$U_1$} & \multicolumn{1}{l|}{\cellcolor[HTML]{FFFFFF}$\mathbf{h}_1^{j}(\mathbbmtt{1})$} & \multicolumn{1}{l|}{\cellcolor[HTML]{96FFFB}$\mathbf{h}_1^{j}(\mathbbmtt{2})$} & \multicolumn{1}{l|}{\cellcolor[HTML]{FFFFFF}$\mathbf{h}_1^{j}(\mathbbmtt{1})$} & \multicolumn{1}{l|}{\cellcolor[HTML]{FFFFFF}$\mathbf{h}_1^{j}(\mathbbmtt{1})$} & \multicolumn{1}{l|}{\cellcolor[HTML]{FFFFFF}$\mathbf{h}_1^{j}(\mathbbmtt{1})$} & \multicolumn{1}{l|}{\cellcolor[HTML]{FFFFFF}$\mathbf{h}_1^{j}(\mathbbmtt{1})$} & \multicolumn{1}{l|}{\cellcolor[HTML]{FFFFFF}$\mathbf{h}_1^{j}(\mathbbmtt{1})$} \\ \hhline{~|-------|}
		\rowcolor[HTML]{FFFFFF} 
		\multicolumn{1}{l|}{\cellcolor[HTML]{FFFFFF}$U_2$} & \multicolumn{1}{l|}{\cellcolor[HTML]{FFFFFF}$\mathbf{h}_2^{j}(\mathbbmtt{1})$} & \multicolumn{1}{l|}{\cellcolor[HTML]{FFFFFF}$\mathbf{h}_2^{j}(\mathbbmtt{1})$} & \multicolumn{1}{l|}{\cellcolor[HTML]{96FFFB}$\mathbf{h}_2^{j}(\mathbbmtt{2})$} & \multicolumn{1}{l|}{\cellcolor[HTML]{FFFFFF}$\mathbf{h}_2^{j}(\mathbbmtt{1})$} & \multicolumn{1}{l|}{\cellcolor[HTML]{FFFFFF}$\mathbf{h}_2^{j}(\mathbbmtt{1})$} & \multicolumn{1}{l|}{\cellcolor[HTML]{FFFFFF}$\mathbf{h}_2^{j}(\mathbbmtt{1})$} & \multicolumn{1}{l|}{\cellcolor[HTML]{FFFFFF}$\mathbf{h}_2^{j}(\mathbbmtt{1})$} \\ \hhline{~|-------|}
		\rowcolor[HTML]{FFFFFF} 
		\multicolumn{1}{l|}{\cellcolor[HTML]{FFFFFF}$U_3$} & \multicolumn{1}{l|}{\cellcolor[HTML]{FFFFFF}$\mathbf{h}_3^{j}(\mathbbmtt{1})$} & \multicolumn{1}{l|}{\cellcolor[HTML]{FFFFFF}$\mathbf{h}_3^{j}(\mathbbmtt{1})$} & \multicolumn{1}{l|}{\cellcolor[HTML]{FFFFFF}$\mathbf{h}_3^{j}(\mathbbmtt{1})$} & \multicolumn{1}{l|}{\cellcolor[HTML]{96FFFB}$\mathbf{h}_3^{j}(\mathbbmtt{2})$} & \multicolumn{1}{l|}{\cellcolor[HTML]{FFFFFF}$\mathbf{h}_3^{j}(\mathbbmtt{1})$} & \multicolumn{1}{l|}{\cellcolor[HTML]{FFFFFF}$\mathbf{h}_3^{j}(\mathbbmtt{1})$} & \multicolumn{1}{l|}{\cellcolor[HTML]{FFFFFF}$\mathbf{h}_3^{j}(\mathbbmtt{1})$} \\ \hhline{~|-------|}
		\rowcolor[HTML]{FFFFFF} 
		\multicolumn{1}{l|}{\cellcolor[HTML]{FFFFFF}$U_4$} & \multicolumn{1}{l|}{\cellcolor[HTML]{FFFFFF}$\mathbf{h}_4^{j}(\mathbbmtt{1})$} & \multicolumn{1}{l|}{\cellcolor[HTML]{FFFFFF}$\mathbf{h}_4^{j}(\mathbbmtt{1})$} & \multicolumn{1}{l|}{\cellcolor[HTML]{FFFFFF}$\mathbf{h}_4^{j}(\mathbbmtt{1})$} & \multicolumn{1}{l|}{\cellcolor[HTML]{FFFFFF}$\mathbf{h}_4^{j}(\mathbbmtt{1})$} & \multicolumn{1}{l|}{\cellcolor[HTML]{96FFFB}$\mathbf{h}_4^{j}(\mathbbmtt{2})$} & \multicolumn{1}{l|}{\cellcolor[HTML]{FFFFFF}$\mathbf{h}_4^{j}(\mathbbmtt{1})$} & \multicolumn{1}{l|}{\cellcolor[HTML]{FFFFFF}$\mathbf{h}_4^{j}(\mathbbmtt{1})$} \\ \hhline{~|-------|} 
		\rowcolor[HTML]{FFFFFF} 
		\multicolumn{1}{l|}{\cellcolor[HTML]{FFFFFF}$U_5$} & \multicolumn{1}{l|}{\cellcolor[HTML]{FFFFFF}$\mathbf{h}_5^{j}(\mathbbmtt{1})$} & \multicolumn{1}{l|}{\cellcolor[HTML]{FFFFFF}$\mathbf{h}_5^{j}(\mathbbmtt{1})$} & \multicolumn{1}{l|}{\cellcolor[HTML]{FFFFFF}$\mathbf{h}_5^{j}(\mathbbmtt{1})$} & \multicolumn{1}{l|}{\cellcolor[HTML]{FFFFFF}$\mathbf{h}_5^{j}(\mathbbmtt{1})$} & \multicolumn{1}{l|}{\cellcolor[HTML]{FFFFFF}$\mathbf{h}_5^{j}(\mathbbmtt{1})$} & \multicolumn{1}{l|}{\cellcolor[HTML]{96FFFB}$\mathbf{h}_5^{j}(\mathbbmtt{2})$} & \multicolumn{1}{l|}{\cellcolor[HTML]{FFFFFF}$\mathbf{h}_5^{j}(\mathbbmtt{1})$} \\ \hhline{~|-------|}
		\multicolumn{1}{l|}{$U_6$}                         & \multicolumn{1}{l|}{$\mathbf{h}_6^{j}(\mathbbmtt{1})$}                         & \multicolumn{1}{l|}{$\mathbf{h}_6^{j}(\mathbbmtt{1})$}                         & \multicolumn{1}{l|}{$\mathbf{h}_6^{j}(\mathbbmtt{1})$}                         & \multicolumn{1}{l|}{$\mathbf{h}_6^{j}(\mathbbmtt{1})$}                         & \multicolumn{1}{l|}{$\mathbf{h}_6^{j}(\mathbbmtt{1})$}                         & \multicolumn{1}{l|}{$\mathbf{h}_6^{j}(\mathbbmtt{1})$}                         & \multicolumn{1}{l|}{\cellcolor[HTML]{96FFFB}$\mathbf{h}_6^{j}(\mathbbmtt{2})$} \\ \hhline{~|-------|} 
	\end{tabular}
\end{table}

\begin{table}[ht]
	\centering
	\caption{Beamforming vectors of BIA}
	\label{fig:biabfvectors}
	\renewcommand\arraystretch{1.5} 
	\begin{tabular}{l|l|}
		\cline{2-2}
		$U_1$ & $\bm{V}_1=\left[\ \mathbf{I}\ \mathbf{I}\ \mathbf{0}\ \mathbf{0}\ \mathbf{0}\ \mathbf{0}\ \mathbf{0}\ \right]^{\text{T}}$ \\ \cline{2-2} 
		$U_2$ & $\bm{V}_2=\left[\ \mathbf{I}\ \mathbf{0}\ \mathbf{I}\ \mathbf{0}\ \mathbf{0}\ \mathbf{0}\ \mathbf{0}\ \right]^{\text{T}}$ \\ \cline{2-2} 
		$U_3$ & $\bm{V}_3=\left[\ \mathbf{I}\ \mathbf{0}\ \mathbf{0}\ \mathbf{I}\ \mathbf{0}\ \mathbf{0}\ \mathbf{0}\ \right]^{\text{T}}$ \\ \cline{2-2} 
		$U_4$ & $\bm{V}_4=\left[\ \mathbf{I}\ \mathbf{0}\ \mathbf{0}\ \mathbf{0}\ \mathbf{I}\ \mathbf{0}\ \mathbf{0}\ \right]^{\text{T}}$ \\ \cline{2-2} 
		$U_5$ & $\bm{V}_5=\left[\ \mathbf{I}\ \mathbf{0}\ \mathbf{0}\ \mathbf{0}\ \mathbf{0}\ \mathbf{I}\ \mathbf{0}\ \right]^{\text{T}}$ \\ \cline{2-2} 
		$U_6$ & $\bm{V}_6=\left[\ \mathbf{I}\ \mathbf{0}\ \mathbf{0}\ \mathbf{0}\ \mathbf{0}\ \mathbf{0}\ \mathbf{I}\ \right]^{\text{T}}$ \\ \cline{2-2} 
	\end{tabular}
\end{table}
\renewcommand{\arraystretch}{1}

This subsection introduces the design of the RA-based BIA scheme \cite{Gou2011AimingPerfectlyintheDarkBIA} in our considered system. BIA achieves interference-free transmission without the need for CSIT in multi-user systems, but it requires specific channel conditions, termed channel patterns. By using RA to artificially change the channel state, we can reconstruct the required channel patterns, known as 'supersymbol'. Incorporating supersymbols, BIA proposes transmit beamforming vectors and decoding matrices without requiring CSIT, thereby fully eliminating multi-user interference. According to the previous design \cite{Gou2011AimingPerfectlyintheDarkBIA}, BIA achieves $N_t K({N_t-1})^{K-1}$ degrees of freedom over a supersymbol with the length of ${(N_t-1)}^K+K({N_t-1})^{K-1}$ in a K-user $N_t\times 1$ channel. Thus the average multiplexing gain of BIA is $\frac{N_t K({N_t-1})^{K-1}}{{(N_t-1)}^K+K({N_t-1})^{K-1}}=\frac{N_t K}{N_t+K-1}$. However, the supersymbol duration scales exponentially as ${N_t}^K$, rendering BIA impractical for dynamic IoT deployments with large user populations.

As an example in a six-user $2\times 1$ channel, the supersymbols and transmitting beamforming vectors $\bm{V}_k \in \mathbb{C}^{14\times 2}$ on the $j$-th tone are shown in Table~\ref{fig:biasupersymbol} and Table~\ref{fig:biabfvectors}
, respectively, where $U_k$ represents user-$k$ and `slot' represents the signal duration. Note that in Table~\ref{fig:biabfvectors}, $\mathbf{I}$ denotes the $2 \times 2$ identity matrix, $\mathbf{0}$ denotes the $2 \times 2$ all-zero matrix.
In this paper, we consider independent BIA design at each subcarrier.

Without loss of generality, consider the $j$-th tone at the first user, the received signal can be expressed as
\begin{equation}
\begin{aligned}
\mathbf{Y}_1^j &= \mathcal{H}_1^j\Big(\bm{V}_1\mathbf{u}_1^j+\bm{V}_2\mathbf{u}_2^j+\dots +\bm{V}_6\mathbf{u}_6^j\Big) +\mathbf{Z}_{1}^{j}
\\
&= \left[\begin{matrix}
\mathbf{h}_1^{j}(\mathbbmtt{1})\\\mathbf{h}_1^{j}(\mathbbmtt{2})\\\mathbf{0}\\\mathbf{0}\\\mathbf{0}\\\mathbf{0}\\\mathbf{0}
\end{matrix}\right]\mathbf{u}_1^j
 + \dots +\left[\begin{matrix}
\mathbf{h}_1^{j}(\mathbbmtt{1})\\\mathbf{0}\\\mathbf{0}\\\mathbf{0}\\\mathbf{0}\\\mathbf{0}\\\mathbf{h}_1^{j}(\mathbbmtt{1})
\end{matrix}\right]\mathbf{u}_6^j
+\mathbf{Z}_{1}^{j},
\end{aligned}
\label{BIAeq1}
\end{equation} 
{where $\mathbf{u}_k^j \in \mathbb{C}^{2\times 1}$ is signal vector for user-$k$ on the $j$-th tone and }

\begin{equation}
\mathcal{H}_1^j= blkdiag\left\{
	\mathbf{h}_1^{j}(\mathbbmtt{1}),\mathbf{h}_1^{j}(\mathbbmtt{2}),\mathbf{h}_1^{j}(\mathbbmtt{1}),\cdots,\mathbf{h}_1^{j}(\mathbbmtt{1})\right\}
.
\label{BIAeq2}	
\end{equation}

Further, an interference cancellation (IC) decoding matrix
\begin{equation}\label{P1-decMat}
\mathbf{P}_1=\left[\begin{matrix}\frac{1}{\sqrt6}& 0& -\frac{1}{\sqrt6}& -\frac{1}{\sqrt6}& -\frac{1}{\sqrt6}& -\frac{1}{\sqrt6}& -\frac{1}{\sqrt6}\\0&1& 0& 0& 0& 0& 0 \\\end{matrix}\right]
\end{equation}
is utilized, hence we have
\begin{equation}
	\label{eq:BIAdecoding}
\mathbf{P}_1\mathbf{Y}_1^j=\bm{H}_1^{j}\mathbf{u}_1^j +\widetilde{\mathbf{Z}}_1^j
,
\end{equation}
where 
\begin{equation}
\bm{H}_1^{j}=\left[\begin{matrix}\frac{1}{\sqrt6}\mathbf{h}_1^{j}(\mathbbmtt{1})\\ \mathbf{h}_1^{j}(\mathbbmtt{2})\\\end{matrix}\right]
\end{equation}
and $\widetilde{\mathbf{Z}}_1^j$ is the additive white Gaussian noise (AWGN) vector with the same mean and variance of $\mathbf{Z}_1^j$. Consequently, an equivalent $2\times 2$ MIMO channel is achieved and traditional MIMO decoders can be readily introduced, e.g., ZF decoder and MMSE decoder.

\begin{figure*}[ht]
	\centering
	\includegraphics[width=6.9in]{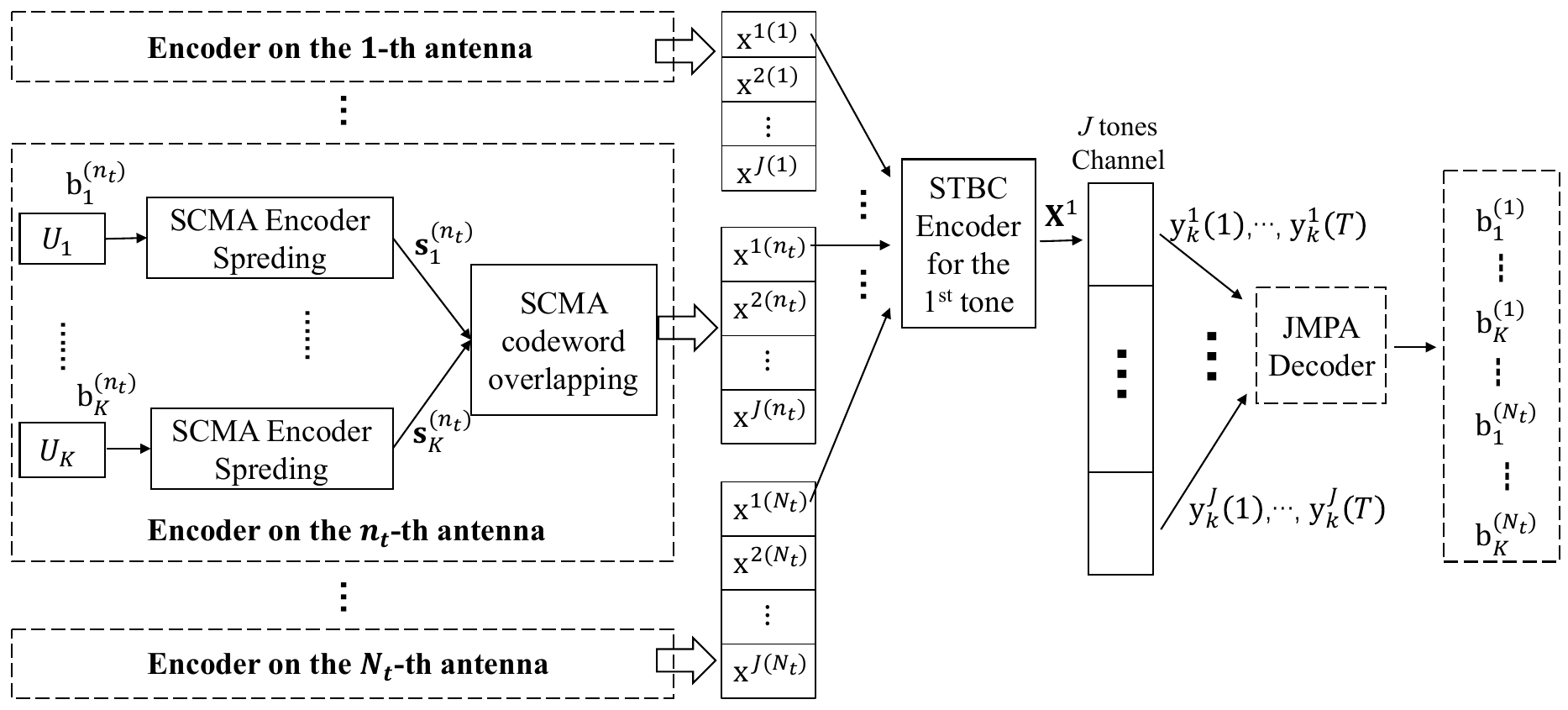}
	\caption{Transceiver diagram of STBC-SCMA scheme with JMPA decoders.}
	\label{SCMA_model}
\end{figure*}

\subsection{STBC-based SCMA }
\label{SCMA_scheme}

\begin{figure*}[ht]
	\centering
	\includegraphics[width=6.8in]{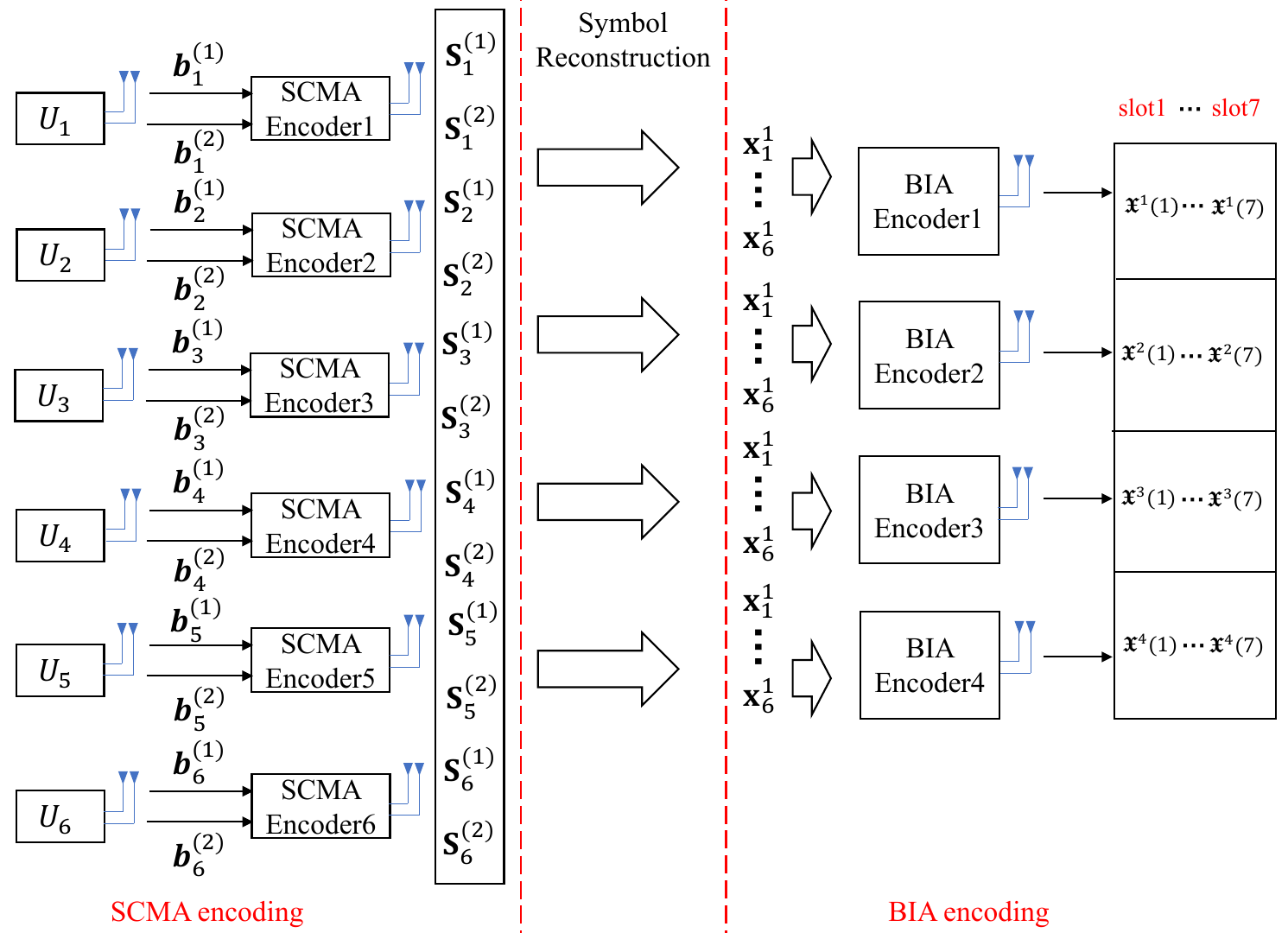}
	\caption{Encoding diagram of the proposed SBMA scheme.}
	\label{fig:encoding}
\end{figure*}

\begin{figure}[ht]
	\centering
	\includegraphics[width=2.5in]{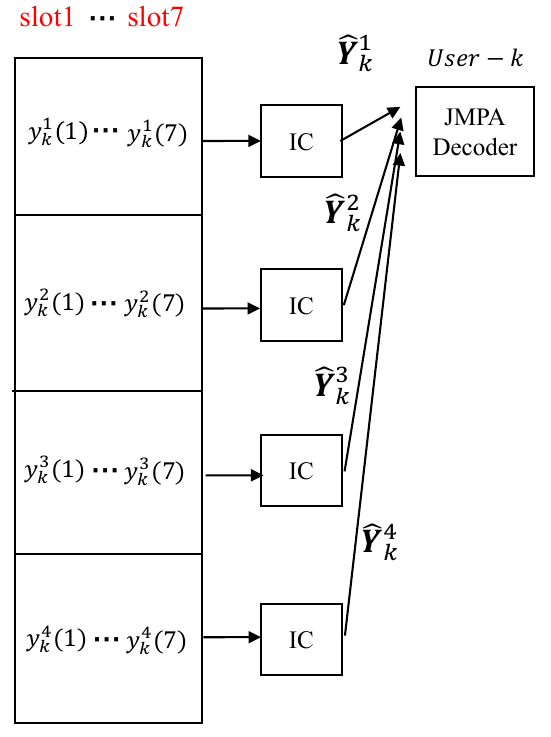}
	\caption{Decoding diagram of the proposed SBMA scheme with JMPA decoders.}
	\label{fig:decoding}
\end{figure}

MIMO techniques are commonly used for spatial multiplexing and diversity. However, in MISO-SCMA systems, MIMO-based spatial multiplexing increases decoding complexity and degrades error performance. Thus, the STBC-SCMA scheme, which uses multiple transmit antennas for spatial diversity gain, is preferred \cite{Pan2019STBC}.

Fig.~\ref{SCMA_model} illustrates the transceiver diagram of the STBC-SCMA scheme, consisting of the encoding process and the decoder design for the $k$-th user. To simplify the illustration, we display only the encoding streams on the first tone. Note that the JMPA is considered to decode all transmitted signals in this figure.
In this model, $\mathrm{b}_k^{(n_{t})}$ denotes the binary input for the $k$-th user at the $n_{t}$-th antenna. Applying the SCMA encoder spreading, $\mathrm{b}_k^{(n_{t})}$ is mapped to a complex codeword $\mathbf{s}_k^{(n_t)}=[s_k^{1(n_t)},\ \cdots ,\ s_k^{J(n_t)}]^{\text{T}}$ for user-$k$ from the $n_t$-th antenna over $J$ subcarriers. Denotes an indicator matrix as $\mathbf{F}=\left[\begin{matrix}\mathbf{f}_1&\cdots&\mathbf{f}_K\\\end{matrix}\right]$, where '1' in $\mathbf{f}_k$ denotes the positions of non-zero elements in $\mathbf{s}_k^{(n_t)}$. Given that $K=6$ and $J=4$, the indicator matrix is then denoted as
\begin{equation}
	\label{Eq:SCMA_F}
	\begin{aligned}
		\mathbf{F}&=\left[\begin{matrix}\mathbf{f}_1&\cdots&\mathbf{f}_K\\\end{matrix}\right]\\
		&=\left[
		\begin{matrix}
			1&1&1&0&0&0\\
			1&0&0&1&1&0\\
			0&1&0&1&0&1\\
			0&0&1&0&1&1\\
		\end{matrix}
		\right],
	\end{aligned}
\end{equation}
where $\mathbf{f}_1$ denotes that $\mathbf{s}_1^{(n_t)}$ contains non-zero symbols $s_1^{1(n_t)}$ and $s_1^{2(n_t)}$, leaving $s_1^{3(n_t)}=s_1^{4(n_t)}=0$. As an example in the above design, the binary number $\mathrm{b}_1^{(n_{t})}=0$ is encoded as $\mathbf{s}_1^{(n_t)}=[-0.6351+0.4615j, 0.1392-0.1759j, 0, 0]^{\text{T}}$, whereas $\mathrm{b}_1^{(n_{t})}=1$ is encoded as $\mathbf{s}_1^{(n_t)}=[0.1815-0.1318j, 0.4873-0.6156j, 0, 0]^{\text{T}}$. Note that the indicator matrix design influences the decoding complexity and accuracy; the one adopted above is designed to yield a moderate performance and computation consumption. Here, all users are using different sets of tones.

After the SCMA encoder spreading, the SCMA codewords $\mathbf{s}_k^{(n_t)}$ are superposed on each tone, resulting in  $\mathbf{x}^{(n_t)}=\sum_{k=1}^{K} \mathbf{s}_k^{(n_t)}=[x^{1(n_t)},\ \cdots,\ x^{J(n_t)}]^{\text{T}}$. This represents the superposed transmitting symbols at the $n_t$-th antenna across all tones. Focusing on the $j$-th tone, the streams $(x^{j(1)},x^{j(2)},\cdots,x^{j(N_t)})$ from the transmit antennas are further processed by an STBC encoder. STBC encodes the streams into $T$ time slots, producing a final transmission matrix $\mathbf{X}^j \in \mathbb{C}^{N_t \times T}$.
Our system uses the conventional Alamouti code as the STBC code for the $2\times 1$ channel, thus the size of $\mathbf{X}^j$ is $2\times 2$. Following the design of Alamouti, the transmitting matrix is formed as
\begin{equation}
\label{Eq:alamouti}
\mathbf{X}^j=\left[\begin{matrix}
	x^{j(1)}&-x^{j(2)*}\\
	x^{j(2)}&x^{j(1)*}
\end{matrix}\right].
\end{equation}

Denotes $y_k^j(t)$ as the received signal of the $k$-th user on the $j$-th tone at the $t$-th slot. According to the Alamouti design in (\ref{Eq:alamouti}), the overall received signals of the $k$-th user on the $j$-th tone are
\begin{equation}
	\label{Eq:SCMArate}
\begin{aligned}
\left[\begin{matrix}y_k^j\left(1\right)\\y_k^{j\ast}\left(2\right)\\\end{matrix}\right]&=
\left[\begin{matrix}h_k^{j(1)}&h_k^{j(2)}\\h_k^{j(2)\ast}&-h_k^{j(1)\ast}\\\end{matrix}\right]
\left[\begin{matrix}x^{j(1)}\\x^{j(2)}\\\end{matrix}\right]
+\left[\begin{matrix}z_k^{j}(1)\\z_k^{j\ast}(2)\\\end{matrix}\right] \\
&=\left[\begin{matrix}h_k^{j(1)}&h_k^{j(2)}\\h_k^{j(2)\ast}&-h_k^{j(1)\ast}\\\end{matrix}\right]
\left[\begin{matrix}\sum_{i=1}^6 s_i^{j(1)}\\\sum_{i=1}^6 s_i^{j(2)}\\\end{matrix}\right]
+\left[\begin{matrix}z_k^{j}(1)\\z_k^{j\ast}(2)\\\end{matrix}\right],
\end{aligned}
\end{equation}
where $h_k^{j(n_t)}$ denotes the CSI from the $n_t$-th antenna of the BS to the $k$-th user on the $j$-th subcarrier, and $z_k^{j}(t)$ denotes the corresponding AWGN at the $t$-th time slot.

Note that the STBC-SCMA decoder design has two options: a low-complexity two-stage decoder and a performance-optimized JMPA decoder. The two-stage decoder is to use a traditional linear MIMO decoder first, then to perform MPA. In specific, linear MIMO decoders are used to decode $x^{j(n_t)}$ in (\ref{Eq:SCMArate}), and then MPA is used to further decode the binary input $\mathrm{b}_k^{(n_{t})}$. The JMPA decoder constructs a virtual factor graph and performs MPA decoding according to the received signals directly. Since the decoder designs for our proposed scheme are similar to the STBC-SCMA decoder, we elaborate on the details of the decoders in the next section.
%


\section{Transceiver design of SBMA}
\label{B-SCMA}

BIA enables interference-free transmission without CSIT, appealing to IoT networks with limited feedback. Yet, it requires long supersymbols (e.g., 7 slots for $K=6$, $N_t=2$), imposing challenges on channel coherence. STBC-SCMA integrates space-time block coding into SCMA for spatial diversity in MISO systems. While enhancing reliability, it still preserves SCMA's privacy risks. Since MPA decodes both desired and interference signals, users' data may be exposed. To address these, we propose SBMA, a scheme combining SCMA and BIA for better performance.

Fig. \ref{fig:encoding} and Fig.~\ref{fig:decoding} show the transceiver diagram of the proposed SBMA design when $K=6$, $J=4$, and $N_t=2$, which is composed of the encoding process and the $k$-th user's decoder design. 

\subsection{Encoding design}
\label{SBMAEncoding}

The encoding process can be presented as two steps: SCMA encoding and BIA encoding, as shown in  Fig. \ref{fig:encoding}.

\subsubsection{Step 1 - SCMA encoding}

First, we define a set of superusers $U_k$, where $k = 1, 2, \ldots, K_U$. For simplicity, by default we assume each superuser corresponds to one physical user, i.e., $K_U=K$ and all $\mathbf{b}_k^{(n_t)}$ for $U_k$ are intended for the same user. For each superuser $U_k$, the BS processes a binary vector $\mathbf{b}_k^{(n_t)} \in \mathbb{B}^{L \times 1}$ for each transmit antenna $n_t$, with $L = 6$ in this study.

Unlike generic SCMA, where $L$ streams of $\mathbf{b}_k^{(n_t)}$ are allocated to different users, SBMA allows flexible allocation. Specifically, $L$ data of $\mathbf{b}_k^{(n_t)}$ can be assigned to a single physical user, or distributed over no more than $L$ physical users, which form a superuser. 

let $b_{k}^{(n_t)}\left(l\right)$ be the $l$-th element of $\bm{b}_k^{(n_t)}$, i.e., a one-bit binary variable. Note that the value of $L$ depends on the number of available subcarriers, $J$, and the predetermined degree of resource reuse. More bits and larger $L$ can be used, but the corresponding codebook design would be more complicated.

The SCMA encoders can be defined as a mapping function $\mathfrak{f} :\mathbb{B}\to \bm{\mathcal{C}}_{kl}^{(n_t)}$, where $\bm{\mathcal{C}}_{kl}^{(n_t)}\subset \mathbb{C}^J$ is the codebook of $b_{k}^{(n_t)}\left(l\right)$ and $\lVert \bm{\mathcal{C}}_{kl}^{(n_t)} \rVert=2$. Thus, we have $\mathbf{s}_{kl}^{(n_t)}=\mathfrak{f}(b_{k}^{(n_t)}\left(l\right))$, where the SCMA codeword $\mathbf{s}_{kl}^{(n_t)}\in \bm{\mathcal{C}}_{kl}^{(n_t)}$ is a sparse vector. 

After the SCMA encoding, $b_{k}^{(n_t)}\left(l\right)$ is mapped into a codeword $\mathbf{s}_{kl}^{(n_t)}$ and superposed over $L$ bits, producing the transmitting symbol vector $\mathbf{S}_k^{(n_t)}=\sum_{l=1}^L \mathbf{s}_{kl}^{(n_t)}\in \mathbb{C}^J $. Denotes $\text{S}_k^{j(n_t)}$ as the $j$-th element in $\mathbf{S}_k^{(n_t)}$. 

Further, we need to reconstruct all SCMA symbol vectors. Focusing on each tone, we denote $\mathbf{x}_k^j=\left[\text{S}_k^{j(1)} \ \cdots \text{S}_k^{j(N_t)}\right]^{\text{T}}$ as the corresponding transmitting vector on the $j$-th tone. In this paper we assume $N_t=2$, thus $\mathbf{x}_k^j=\left[\text{S}_k^{j(1)} \ \text{S}_k^{j(2)}\right]^{\text{T}}$.

\subsubsection{Step 2 - BIA encoding}
BIA encoding is applied at each tone by employing symbol extension in the time domain. This section uses the same BIA supersymbol and beamforming vectors as shown in Sect.~\ref{modelBIA}. Focusing on the $j$-th tone, the transmitting vector after reconfigurable-antenna-based BIA can be derived as
\begin{equation}
\mathbf{\mathfrak{X}}^j=\left[\begin{matrix}
\mathbf{\mathfrak{X}}^j(1)\\\mathbf{\mathfrak{X}}^j(2)\\\mathbf{\mathfrak{X}}^j(3)\\\mathbf{\mathfrak{X}}^j(4)\\\mathbf{\mathfrak{X}}^j(5)\\\mathbf{\mathfrak{X}}^j(6)\\\mathbf{\mathfrak{X}}^j(7)
\end{matrix}
\right]=
\left[\begin{matrix} \mathbf{x}_1^{j}+\mathbf{x}_2^{j}+\mathbf{x}_3^{j}+\mathbf{x}_4^{j}+\mathbf{x}_5^{j}+\mathbf{x}_6^{j}\\\mathbf{x}_1^{j}\\\mathbf{x}_2^{j}\\\mathbf{x}_3^{j}\\\mathbf{x}_4^{j}\\\mathbf{x}_5^{j}\\\mathbf{x}_6^{j} \end{matrix}\right],
\end{equation}
where $\mathbf{\mathfrak{X}}^{j}\left({t}\right)\in\mathbb{C}^{N_t\times1}$ denotes the transmitting vector from the BS on the $j$-th tone at the $t$-th slot.

After the encoding, the received signals at the first user can be formulated as
\begin{equation}
	\bm{Y}_{1}=\left[\begin{matrix}\left(\bm{y}_1^{1}\right)^{\text{T}}&\cdots&\left(\bm{y}_1^{J}\right)^{\text{T}}\\\end{matrix}\right]^{\text{T}},
\end{equation}
where the received signals on the $j$-th tone over 7 slots are
\begin{equation}
	\begin{aligned}
		\bm{y}_1^{j}=&
		{\left[\begin{matrix}
				y_1^{j}\left(1\right)&y_1^{j}\left(2\right)&\cdots&y_1^{j}\left(7\right)
			\end{matrix}\right]}^{\text{T}}\\
		=&\left[\begin{matrix}\mathbf{h}_1^{j}(\mathbbmtt{1})&\mathbf{0}&\cdots&\mathbf{0}\\\mathbf{0}&\mathbf{h}_1^{j}(\mathbbmtt{2})&\cdots&\mathbf{0}\\\vdots&\vdots&\ddots&\mathbf{0}\\\mathbf{0}&\mathbf{0}&\mathbf{0}&\mathbf{h}_1^{j}(\mathbbmtt{1})\\\end{matrix}\right]\left[\begin{matrix}\mathbf{\mathfrak{X}}^j(1)\\\vdots\\\mathbf{\mathfrak{X}}^j(7)\\\end{matrix}\right]+\bm{Z}_1^{j},
	\end{aligned}
\end{equation}
where $\bm{Z}_1^{j}=\left[\begin{matrix}z_1^{j}\left(1\right)&\cdots&z_1^{j}\left(7\right)\\\end{matrix}\right]^{\text{T}}$ is the AWGN vector over 7 slots and $\mathbf{0}$ denotes a all-zero matrix with the size of $1\times 2$.

\subsubsection{Notation for codebook design in SBMA}\label{NotCoBook}
In SCMA, distinct codebooks are utilized for multiple streams, enabling the MPA to decode them effectively. 
In SBMA, different streams within each superuser utilize distinct codebooks to facilitate effective MPA decoding, similar to SCMA. However, because BIA eliminates user interference and assumes that the channels across different antennas are uncorrelated, the same codebook can be reused across various superusers and transmit antennas without confusion. This shared codebook architecture simplifies system design and reduces the complexity of codebook management. So $\bm{\mathcal{C}}_{kl}^{(n_t)}$ can be simplified as $\bm{\mathcal{C}}_l$, which is the codebook for the $l$-th stream and is used uniformly across all superusers and antennas.

\subsection{Two-stage Decoding}

Here, we propose a two-stage decoding design. The first stage involves a traditional linear BIA decoder, which consists of a multi-user interference cancellation (IC) block and a linear MIMO decoder, specifically a Zero-Forcing decoder. This stage effectively eliminates multi-user interference and separates spatial data streams. In the second stage, a conventional log-MPA is employed to decode SCMA signals for each user. 

The two-stage decoder achieves low decoding complexity primarily due to the linear BIA decoder. By employing a linear BIA decoding process, the multi-user interference is eliminated early on, reducing the overall complexity of subsequent signal processing steps, such as the MPA decoding stage. This enables efficient handling of data streams while minimizing the computational burden typically associated with multi-user decoding.

\subsubsection{BIA decoding}
In this step, we apply the BIA decoding for signals over each tone, respectively. The BIA decoding comprises multi-user interference cancellation and signal detection. A multi-user IC decoding matrix $\bm{P}_k$ is firstly used for the $k$-th user at each tone. As shown in \eqref{P1-decMat}, we focus on the first user and illustrate its two-stage decoding design, e.g.,
\begin{equation*}
\bm{P}_1=\left[\begin{matrix}\frac{1}{\sqrt6}&0&-\frac{1}{\sqrt6}&-\frac{1}{\sqrt6}&-\frac{1}{\sqrt6}&-\frac{1}{\sqrt6}&-\frac{1}{\sqrt6}\\0&1&0&0&0&0&0\\\end{matrix}\right]
.
\end{equation*}
Denote ${\hat{\bm{Y}}}_{k}^j$ as the received signal vector after the multi-user IC on the $j$-th tone, hence the received signals over four tones can be formulated as
\begin{equation}
\begin{aligned}
{\hat{\bm{Y}}}_{1}&=\left[\begin{matrix}
	\left(\hat{\bm{Y}}_{1}^1\right)^{\text{T}}&\left(\hat{\bm{Y}}_{1}^2\right)^{\text{T}}&\left(\hat{\bm{Y}}_{1}^3\right)^{\text{T}}&\left(\hat{\bm{Y}}_{1}^4\right)^{\text{T}}&
\end{matrix}\right]^{\text{T}}\\
&={(\bm{I}_4\otimes\bm{P}}_1)\bm{Y}_{1}\\
&=\left[\begin{matrix}\bm{H}_1^{1}&\mathbf{0}&\mathbf{0}&\mathbf{0}\\\mathbf{0}&\bm{H}_1^{2}&\mathbf{0}&\mathbf{0}\\\mathbf{0}&\mathbf{0}&\bm{H}_1^{3}&\mathbf{0}\\\mathbf{0}&\mathbf{0}&\mathbf{0}&\bm{H}_1^{4}\\\end{matrix}\right]\left[\begin{matrix}
\mathbf{x}_{1}^{1}\\\mathbf{x}_{1}^{2}\\\mathbf{x}_{1}^{3}\\\mathbf{x}_{1}^{4}\end{matrix}\right]+\left[\begin{matrix}{\hat{\bm{Z}}}_1^{1}\\{\hat{\bm{Z}}}_1^{2}\\{\hat{\bm{Z}}}_1^{3}\\{\hat{\bm{Z}}}_1^{4}\\\end{matrix}\right]\\
&=\hat{\bm{H}}_1 \mathbf{X}_1+\hat{\bm{Z}}_1,
\end{aligned}
\label{received_sig}
\end{equation}
where $\bm{H}_k^{j}=\left[\begin{matrix}\frac{1}{\sqrt6}{\mathbf{h}_k^{j}(\mathbbmtt{1})}^{\text{T}}&{\mathbf{h}_k^{j}(\mathbbmtt{2})}^{\text{T}}\\\end{matrix}\right]^{\text{T}}\in\mathbb{C}^{2\times2}$ is the corresponding channel matrix for the $k$-th user on the $j$-th tone. Note that in (\ref{received_sig}), $\mathbf{0}$ denotes a all-zero matrix with the size of $2\times 2$. Besides, the corresponding noise vector is
\begin{equation}
	{\hat{\bm{Z}}}_1^{j}=\left[\begin{matrix}\frac{1}{\sqrt6}z_1^{j}\left(1\right)-\frac{1}{\sqrt6}\left(\sum_{t=3}^{7}{z_1^{j}\left(t\right)}\right)\\z_1^{j}\left(2\right)\\\end{matrix}\right]
.
\end{equation}
Since $\bm{H}_1^{j}$ is full-rank, we can decode each symbol of $\mathbf{X}_1$ by zero-forcing. Similarly, we have untied the two spatial data streams on each tone and obtained the estimated vector ${\hat{\mathbf{x}}}_{k}^{j}=\left[\begin{matrix}
	{\hat{\text{x}}}_{k}^{{j}}(1)&{\hat{\text{x}}}_{k}^{j}(2)
\end{matrix}\right]^{\text{T}}$.
\subsubsection{MPA decoding}
\begin{figure}
	\centering
	\includegraphics[width=3.4in]{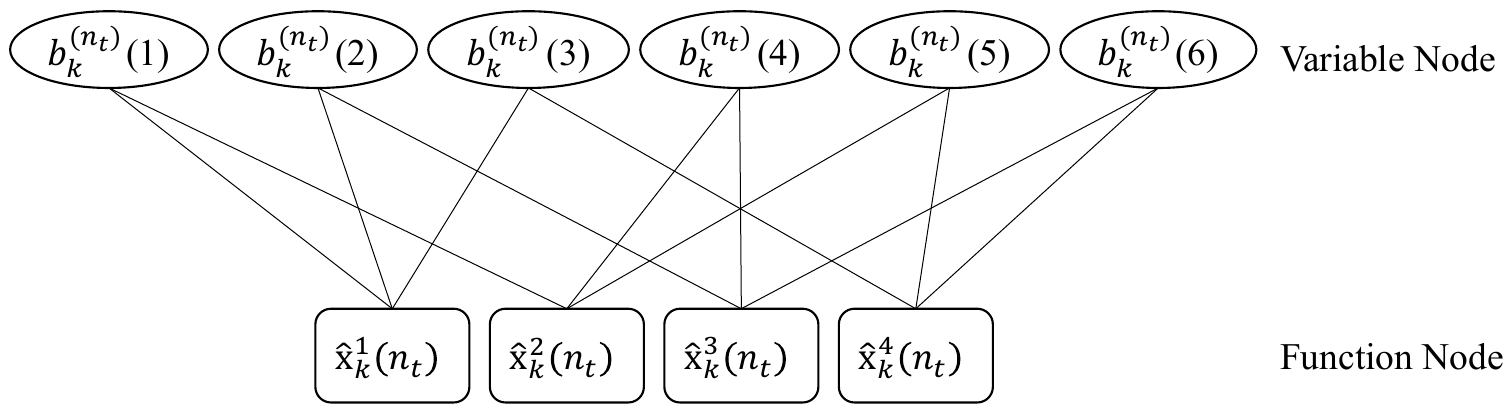}
	\caption{Factor graph for the $k$-th user of the $n_t$-th spatial data stream when $J=4$ and $L=6$.}
	\label{fig:mpa}
\end{figure}

Since the BIA decoder has separated spatial data streams, a simple MPA decoder is further applied to each spatial data stream. Fig. \ref{fig:mpa} shows
the MPA factor graph for the user-$k$ of the $n_t$-th spatial data stream. Since the separated spatial data streams consist solely of superposed SCMA codewords, the MPA decoder used in this design does not require CSI, resulting in a lower computational complexity compared to MPA decoders in conventional SCMA schemes. In this step, we use ${\hat{\text{x}}}_{k}^{{1}}(n_t),{\hat{\text{x}}}_{k}^{2}(n_t),{\hat{\text{x}}}_{k}^{3}(n_t),{\hat{\text{x}}}_{k}^{4}(n_t)$ as function nodes, then we can decode $\bm{b}_{k}^{({n_t})}$ at the $n_t$-th spatial data stream.

\subsection{JMPA decoding}
\begin{figure*}
	\centering
	\includegraphics[width=6.3in]{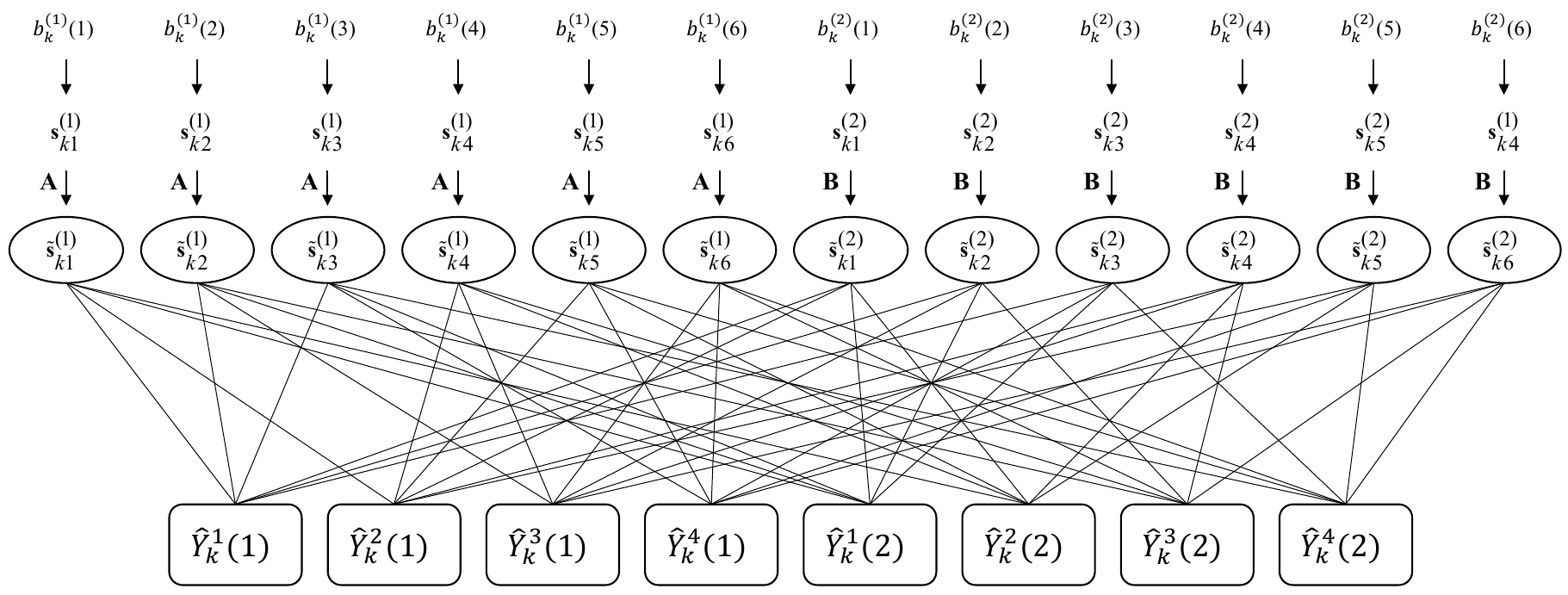}
	\caption{Virtual factor graph for the $k$-th user of the JMPA when $J=4$, $L=6$ and $N_t=2$.}
	\label{fig:decoding2}
\end{figure*} 
Despite the low decoding complexity achieved, the two-stage decoder suffers from limited BER performance due to the misalignment between the traditional linear MIMO decoder and the MPA decoder. Specifically, CSI is employed solely in the first stage, giving no advantage to the MPA decoder in the subsequent stage. To improve the performance further, we propose a JMPA design for SBMA, drawing insights from the conventional JMPA used in MIMO-SCMA \cite{Tang2016JMPA}. As shown in Fig. \ref{fig:encoding}, the designed decoder comprises multi-user interference cancellation blocks coupled with a JMPA decoder.

Different from the two-stage decoding, the received signal for user-k after multi-user IC on the $j$-th tone, i.e., $ {\hat{\bm{Y}}}_{k}^j=\left[\begin{matrix}
	\hat{Y}_{k}^j(1)&\cdots&\hat{Y}_{k}^j(N_t)
\end{matrix}\right]^{\text{T}}$, are treated as function nodes directly then decode all spatial data streams simultaneously. Fig. \ref{fig:decoding2} shows the virtual factor graph for the $k$-th user, which consists of $N_t L$ variable nodes \big( $b_{k}^{(n_t)}\left(l\right)$ \big) and $N_t J$ function nodes \big($\hat{Y}_{k}^j(n_t)$\big). Note that the order of elements in ${\hat{\bm{Y}}}_{k}^j$ is rearranged in Fig. \ref{fig:decoding2}. The virtual indicator matrix is
\begin{equation}
	\mathbf{F}_{v}=\left[\begin{matrix}\mathbf{F}&\mathbf{F}\\\mathbf{F}&\mathbf{F}\\\end{matrix}\right]
,
\end{equation}
where the $l'$-th column of $\mathbf{F}_{v}$ describes the connectivity of all function nodes to the $l'$-th variable node.

Following the encoding scheme described in Sect.~\ref{NotCoBook}, where the binary input $b_{k}^{(n_t)}(l)$ is mapped to an SCMA codeword $\mathbf{s}_{kl}^{(n_t)} \in \bm{\mathcal{C}}_{l}$, we now examine the virtual codeword generation process. As illustrated in Fig. \ref{fig:decoding2}, the $l'$-th virtual variable node produces the virtual codeword $\tilde{\mathbf{s}}_{kl}^{(n_t)} \in \hat{\bm{\mathcal{C}}}_{l'}$ with $l'=(n_t-1)L+l$, through transformation of the original codebook $\bm{\mathcal{C}}_l$ using CSI. 
The virtual codebook $\hat{\bm{\mathcal{C}}}_{l'}$ is constructed as follows. First, we define the CSI vector corresponding to the $l'$-th variable node as:
\begin{equation}
	\hat{\mathbf{h}}^{l'}_k = \left[ \hat{h}^{l',1}_k \ \hat{h}^{l',2}_k \ \cdots \ \hat{h}^{l',N_t J}_k \right]^{\text{T}} \in \mathbb{C}^{N_t J \times 1},
\end{equation}
where $\hat{h}^{l',j'}_k$ denotes the CSI coefficient between the $l'$-th variable node and $j'$-th function node. 
The virtual codebook generation is then performed by:
\begin{equation}
	\hat{\bm{\mathcal{C}}}_{l'} = 
	diag\left(\hat{\mathbf{h}}^{l'}_k\right)
	\left[\begin{matrix}
		\bm{\mathcal{C}}_l \\
		\bm{\mathcal{C}}_l
	\end{matrix}\right].
\end{equation}

This formulation enables codebook reuse across multiple transmit antennas, with the CSI transformation creating distinct virtual codebooks for each variable node.

Furthermore, the construction of the CSI vector $\hat{\mathbf{h}}^{l'}_k$ depends on the CSI matrix $\hat{\bm{H}}_k$ and the virtual factor graph structure. Note that variable nodes associated with the same antenna share the same ${\hat{\mathbf{h}}}^{l'}_k$, and ${\hat{\mathbf{h}}}^{l'}_k$ is determined by the transmit antenna and the corresponding tone.

Given that $N_t=2$, we denote that the first $L$ variable nodes share the CSI vector $\mathbf{A}$, corresponding to the first antenna, while the last $L$ variable nodes share the CSI vector $\mathbf{B}$, corresponding to the second antenna. Thus
\begin{equation}
	{\hat{\mathbf{h}}}^{l'}_k=\left\{
	\begin{aligned}
		\mathbf{A},&\ \  l'\leq L\\
		\mathbf{B},&\ \  L<l'\leq N_t L
	\end{aligned} 
	\right
	..
\end{equation}
To determine $\mathbf{A}$ and $\mathbf{B}$, we need to reconstruct $\hat{\bm{H}}_k$ according to the order of the function nodes. Recall the structure of $\hat{\bm{H}}_k$ in (\ref{received_sig}) and that $\bm{H}_k^{j}=\left[\begin{matrix}\frac{1}{\sqrt6}{\mathbf{h}_k^{j}(\mathbbmtt{1})}^{\text{T}}&{\mathbf{h}_k^{j}(\mathbbmtt{2})}^{\text{T}}\\\end{matrix}\right]^{\text{T}}$, where the channel matrix is related to the receive antenna across two operation modes. In Fig. \ref{fig:decoding2}, the first $J$ function nodes are associated with the receive antenna in the first mode, while the last $J$ function nodes are associated with the second mode. Therefore, $\mathbf{A}$ and $\mathbf{B}$ can be determined as follows:
\begin{equation}
	\begin{aligned}
		&\left[\begin{matrix}\mathbf{A}&\mathbf{B}\\\end{matrix}\right]\\
		&=\left[\begin{matrix}\frac{1}{\sqrt6}{\mathbf{h}_k^{1}(\mathbbmtt{1})}^{\text{T}}&\cdots&\frac{1}{\sqrt6}{\mathbf{h}_k^{4}(\mathbbmtt{1})}^{\text{T}}&{\mathbf{h}_k^{1}(\mathbbmtt{2})}^{\text{T}}&\cdots&{\mathbf{h}_k^{4}(\mathbbmtt{2})}^{\text{T}} \end{matrix}\right]^{\text{T}}
	\end{aligned}
	,
\end{equation}
where $\mathbf{A,B}\in\mathbb{C}^{{N_t J}\times1}$.

With the virtual factor graph and the virtual codebook, we can apply traditional Log-MPA to decode all desired signals.

\section{BER analysis of SBMA}
\label{ErrorAna}

This section derives a closed-form BER expression for SBMA with JMPA decoding in the specific case of $K = 6$, $J = 4$, and $L = 6$, validating the diversity order claimed in Sect.~\ref{Sect.Intro}. The analysis, while computationally intensive for larger systems, serves as a foundation for future generalizations. We assume equal power allocation per bit, with $E_b$ denoting the energy per bit, thus we have
\begin{equation}
	\mathbb{E}\left\{\left|\mathbf{x}_{k}^{j}\right|^2\right\}=\frac{L\times N_t E_b}{J\times N_t}=\frac{3E_b}{2}.
	\label{EbPower}
\end{equation}
Note that there is a '$N_t$' in the denominator of (\ref{EbPower}), because each symbol is transmitted $N_t$ times as required by the BIA encoder. For analysis, we denote $\mathbf{x}_{k}^{j}=\frac{3E_b}{2}\hat{\mathbf{x}}_{k}^{j}$, where $\hat{\mathbf{x}}_{k}^{j}$ is normalized transmitting symbol for user-$k$ on the $j$-th tone. Further we have $\mathbf{X}_k=\frac{3E_b}{2}\mathbf{X}_k^{\star}$, where $\mathbf{X}_k^{\star}=\left[\begin{array}{cccc}
\hat{\mathbf{x}}_{k}^{1}&\hat{\mathbf{x}}_{k}^{2}&\hat{\mathbf{x}}_{k}^{3}&\hat{\mathbf{x}}_{k}^{4}
\end{array} \right]$.
  
The received signal can be detected using maximum likelihood (ML) detection, which provides optimal performance. Alternatively, the message-passing-based algorithm can achieve this performance with enough iterations. Therefore, we can evaluate the error performance of SBMA using ML-based analysis.
According to the received signal, the estimated symbol vector $\hat{\mathbf{X}}_k^{\star}$ is determined by
\begin{equation}
	\hat{\mathbf{X}}_k^{\star}=\arg \min \left\Vert \hat{\bm{Y}}_k-\sqrt{\frac{3E_b}{2N_0}}\hat{\bm{H}}_k\mathbf{X}_k^{\star} \right\Vert_F^2
.
\end{equation}

The conditional pairwise error probability (PEP), defined as the probability of $\mathbf{X}_k^{\star}\to \hat{\mathbf{X}}_k^{\star}$ for fixed channel coefficients, can be expressed as
\begin{equation}
\begin{aligned}
&Pr\left(\mathbf{X}_k^{\star}\to \hat{\mathbf{X}}_k^{\star}|\hat{\bm{H}}_k \right)\\
&=Q\left(
\sqrt{\frac{3E_b}{4 N_0}}\left\Vert \hat{\bm{H}}_k(\mathbf{X}_k^{\star} - \hat{\mathbf{X}}_k^{\star}) \right\Vert_F \right),
\end{aligned}
\label{Eq.PEP}
\end{equation}
where $Q(x)=\frac{1}{\sqrt{2\pi}}\int_{x}^{\infty}e^{-\frac{t^2}{2}}dt$. To evaluate (\ref{Eq.PEP}), we use an upper bound for the Q-function as in \cite{Bao2017}, thus we have
\begin{equation}
\begin{aligned}
&Pr\left(\mathbf{X}_k^{\star}\to \hat{\mathbf{X}}_k^{\star}|\hat{\bm{H}}_k \right)\\
&\leq \sum_{n=1}^{N}a_n \exp\left( -c_n \frac{3E_b}{4 N_0}\left\Vert \hat{\bm{H}}_k(\mathbf{X}_k^{\star} - \hat{\mathbf{X}}_k^{\star}) \right\Vert_F^2 \right),
\end{aligned}
\end{equation}
where $N,a_n,c_n$ are constants. Note that the upper bound tends to the exact value as $N$ increases.

Since the fact that
\begin{equation}
\begin{aligned}
&\left\Vert \hat{\bm{H}}_k(\mathbf{X}_k^{\star} - \hat{\mathbf{X}}_k^{\star}) \right\Vert_F^2\\
&=Tr\left( \hat{\bm{H}}_k(\mathbf{X}_k^{\star} - \hat{\mathbf{X}}_k^{\star}){(\mathbf{X}_k^{\star} - \hat{\mathbf{X}}_k^{\star})}^H \hat{\bm{H}}_k^H \right)\\
&=\sum_{l=1}^{8}\bm{h}_k^l(\mathbf{X}_k^{\star} - \hat{\mathbf{X}}_k^{\star}){(\mathbf{X}_k^{\star} - \hat{\mathbf{X}}_k^{\star})}^H{(\bm{h}_k^l)}^H,
\end{aligned}
\end{equation}
where $\bm{h}_k^l$ is the $l$-th row vector of $\hat{\bm{H}}_k$ and note that the matrix $(\mathbf{X}_k^{\star} - \hat{\mathbf{X}}_k^{\star}){(\mathbf{X}_k^{\star} - \hat{\mathbf{X}}_k^{\star})}^H$ is Hermitian that can be diagonalized by a unitary transformation, thus we have
\begin{equation}
\begin{aligned}
&Pr\left(\mathbf{X}_k^{\star}\to \hat{\mathbf{X}}_k^{\star}|\hat{\bm{H}}_k \right)\\
&\leq \sum_{n=1}^{N}a_n \exp\left( -c_n \frac{3E_b}{4 N_0}\sum_{l=1}^{8} \bm{h}_k^l \mathbf{V}_k \mathbf{\Lambda}_k \mathbf{V}_k^H {(\bm{h}_k^l)}^H \right)\\
&=\sum_{n=1}^{N}a_n \exp\left( -c_n \frac{3E_b}{4 N_0}\sum_{l=1}^{8}\sum_{i=1}^{8} \lambda_k^i \left| \bm{h}_k^l \mathbf{v}_k^i \right|^2 \right),
\end{aligned}
\end{equation}
where $\mathbf{\Lambda}_k=diag(\lambda_k^1,\lambda_k^2,\cdots,\lambda_k^8)$. $\mathbf{V}_k$ is a unitary matrix and $\mathbf{v}_k^i=[\begin{matrix}
v_k^{1,i}&v_k^{2,i}&\cdots&v_k^{8,i}
\end{matrix}]^{\text{T}}$ is the $i$-th column of $\mathbf{V}_k$. 

In this paper, we consider the Rayleigh fading channel, i.e., $\mathbf{h}_k^{j}(m)\sim\mathbb{CN}(\mathbf{0},\mathbf{I})$. Denotes $\left| \bm{h}_k^l \mathbf{v}_k^i \right|^2$ as $\beta_k^{l,i}$, thus the probability density function of the random variable $\beta_k^{l,i}$ is given by $f_{\beta_k^{l,i}}(x)=\frac{1}{4b_k^{l,i}}\exp\left( -\frac{x}{4b_k^{l,i}} \right)$, where $b_k^{l,i}=\sum_{j=2\lceil\frac{l}{2}\rceil-1}^{2\lceil\frac{l}{2}\rceil}T_l|v_k^{j,i}|^2$, where
\begin{equation}
	T_l=\left\{
	\begin{aligned}
	\frac{1}{6}& ,\ \ l\ is\ odd\\
	1& ,\ \ l\ is\ even\\
	\end{aligned}
	\right
..
\end{equation}
Note that $T_l$ comes from the fact that $\bm{H}_k^{j}=\left[\begin{matrix}\frac{1}{\sqrt6}{\mathbf{h}_k^{j}(\mathbbmtt{1})}^{\text{T}}&{\mathbf{h}_k^{j}(\mathbbmtt{2})}^{\text{T}}\\\end{matrix}\right]^{\text{T}}$.

Then the average PEP has
\begin{equation}
\begin{aligned}
&Pr\left(\mathbf{X}_k^{\star}\to \hat{\mathbf{X}}_k^{\star}\right)\\
&\leq\sum_{n=1}^{N}a_n\mathbb{E}_{\hat{\bm{H}}_k}\left[\exp\left(-c_n \frac{3E_b}{4 N_0}\sum_{l=1}^{8}\sum_{i=1}^{8} \lambda_k^i \left| \bm{h}_k^l \mathbf{v}_k^i \right|^2
\right)\right]\\
&=\sum_{n=1}^{N}a_n\prod_{l=1}^{8}\prod_{i=1}^{8}\frac{1}{1+c_n\frac{3E_b}{4 N_0}\lambda_k^i \left(4 b_k^{l,i}\right)}.
\end{aligned}
\label{eq.theroExp}
\end{equation}
Hence, the universal bound of average BER of the system should be a weighted version of the PEP
\begin{equation}
P_B=\frac{1}{6}\frac{1}{2^{L\times N_t}}\sum_{k=1}^{6}\sum_{i=1}^{2^{L\times N_t}}\sum_{j=1,j\neq i}^{2^{L\times N_t}}n_{\text{E}}(\mathbf{X}_k^{\star i}, \mathbf{X}_k^{\star j}) Pr(\mathbf{X}_k^{\star i}\to \mathbf{X}_k^{\star j}),
\end{equation}
where $n_{\text{E}}(\mathbf{X}_k^{\star i}, \mathbf{X}_k^{\star j})$ is the number of bits in which $\mathbf{X}_k^{\star i}$ differs from $\mathbf{X}_k^{\star j}$.

Denote $r$ as the rank of $(\mathbf{X}_k^{\star i}-\mathbf{X}_k^{\star j})$, the minimum $r$ will provide the highest contribution to $Pr(\mathbf{X}_k^{\star i}\to \mathbf{X}_k^{\star j})$, thus the diversity order is obtained by considering the virtue factor graph in Fig. \ref{fig:decoding2} as
\begin{equation}
	div = \min r = 4.
\end{equation}
Note the diversity order in SBMA is the same as that of STBC-SCMA \cite{Pan2019STBC}, as both schemes benefit from spatial and tone diversity through their encoding structures.

Note that the BER analysis for the toy system presented in this section is inspired by \cite{Bao2017}. However, this approach becomes computationally prohibitive as the number of users and/or the size of the codebook increases. A more general BER analysis will be addressed in our future work.

\section{Gain analysis and decoding complexity}
\label{Sect.GainAna}

This section discusses the diversity and multiplexing gain of BIA, STBC-SCMA, and SBMA. Besides, the computational complexity of decoders is analyzed. 

Unlike the specific $N_t$, $K$, $K_U$, $J$, and $L$ considered in Sect.~\ref{ErrorAna}, the analysis in this section is applicable to both the setting of  $N_t=2$, $K=K_U=6$, $J=4$, and $L=6$, and more general configurations of these parameters. The overall comparison is presented in Table~\ref{Tb:OverallComp}. Since ZF decoders exhibit inferior performance, the subsequent analyses focus on ML and JMPA-related decoders unless otherwise stated.

\subsection{Diversity gain}
In BIA, each user accesses an equivalent interference-free $N_t \times N_t$ MIMO channel, thereby achieving a diversity order of $N_t$ \cite{Gou2011AimingPerfectlyintheDarkBIA}.

In both SBMA and STBC-SCMA, SCMA encoding facilitates diversity across multiple tones, with each user allocated $\lVert \mathbf{f}_k \rVert$ tones as determined by the indicator matrix. Furthermore, BIA encoding in SBMA and STBC in STBC-SCMA contribute spatial diversity through the $N_t$ transmit antennas. Consequently, both systems attain a diversity order of $\lVert \mathbf{f}_k \rVert \times N_t$.
	
For the specific system configuration outlined in Section~\ref{ErrorAna}, BIA yields a diversity order of 2, whereas STBC-SCMA and SBMA achieve a diversity order of 4.

\subsection{Multiplexing gain}

According to the findings in the previous study \cite{Gou2011AimingPerfectlyintheDarkBIA}, the BIA scheme detailed in Sect.~\ref{modelBIA} achieves $N_t K({N_t-1})^{K-1} J$ degrees of freedom across ${(N_t-1)}^K+K({N_t-1})^{K-1}$ time slots. As a result, the achievable average multiplexing gain is calculated as $\frac{N_t K({N_t-1})^{K-1} J}{{(N_t-1)}^K+K({N_t-1})^{K-1}}=\frac{N_t K J}{N_t+K-1}$.

In SCMA schemes, the multi-dimensional coding mechanism does not yield additional degrees of freedom, resulting in a maximum multiplexing gain of $J$. Similarly, the STBC-SCMA introduces space-time coding to leverage diversity gain from transmit antennas but does not provide extra multiplexing gain, maintaining the same multiplexing gain of $J$.

By contrast, in the SBMA scheme, the BIA encoding effectively mitigates user interferences, thereby offering $N_t K_U({N_t-1})^{K_U-1}$ degrees of freedom across ${(N_t-1)}^{K_U}+K_U({N_t-1})^{K_U-1}$ time slots. Subsequently, the SCMA encoding achieves $J$ degrees of freedom over $J$ tones. Consequently, the multiplexing gain of SBMA equals $\frac{N_t K_U({N_t-1})^{K_U-1} J}{{(N_t-1)}^{K_U}+K_U({N_t-1})^{K_U-1}}=\frac{N_t K_U J}{N_t+K_U-1}$, equal to that of the BIA scheme.

For the specific system configuration detailed in Section~\ref{ErrorAna}, STBC-SCMA provides a multiplexing gain of 4. In contrast, both BIA and SBMA attain a multiplexing gain of $\frac{48}{7}$, given that $K = K_U = 6$.

\subsection{Decoding complexity}
\label{DecodeComplexity}
We then discuss the computational complexity of the two decoders that have been proposed. As per our understanding, the complexity of a ZF decoder is $\mathcal{O}\left( N_t^3 \right)$, whereas the complexity of a conventional MPA is $\mathcal{O}\left( K\lVert \bm{\mathcal{C}}_{l}\rVert ^D\right)$. Here, $D$ corresponds to the number of assumed users on each tone \cite{Yuan2018ALCEMSCMADetector}. For instance, if the indicator matrix (\ref{Eq:SCMA_F}) is used, then $D=3$. 

Accordingly, the proposed two-stage decoder has a complexity of $\mathcal{O}\left(N_t^3+N_t L\lVert \bm{\mathcal{C}}_{l}\rVert^D\right)$. On the other hand, the JMPA has a larger factor graph and a complexity of $\mathcal{O}\left(N_t L\left(\lVert \bm{\mathcal{C}}_{l}\rVert\right) ^{DN_t}\right)$. Since D is usually larger than 2 in SCMA, the JMPA suffers from a higher computational complexity than the two-stage decoder.

\section{Simulation and Discussion}
\label{Simu}

\begin{table*}[]
	\centering	
	\caption{Gain and decoding comparison on three schemes}
	\label{Tb:OverallComp}
	\renewcommand\arraystretch{1.5}	
	\begin{tabular}{ccccl}
		\cline{1-4}
		\toprule[1.5pt]
		& Diversity order                         & Multiplexing gain         & Decoding complexity                                                                                    &  \\ \cline{1-4}
		BIA (ZF)       &  1                                    & $\frac{N_t K J}{N_t+K-1}$                        & $\mathcal{O}\left( N_t^3 \right)$                                                                      &  \\
		STBC-SCMA (ZF-MPA) &  $N_t$                                   & $J$                        & \multirow{2}{*}{$\mathcal{O}\left(N_t^3+N_t L\lVert \bm{\mathcal{C}}_{l}\rVert^D\right)$}              &  \\
		SBMA (ZF-MPA)      &  1                                     & $\frac{N_t K_U J}{N_t+K_U-1}$                        &                                                                                                        &  \\ \cline{1-4}
& & & & \\
		BIA (ML)            & $N_t$                                  & $\frac{N_t K J}{N_t+K-1}$ & -                                                                                                      &  \\
		STBC-SCMA (JMPA)      & $\lVert \mathbf{f}_k \rVert\times N_t$ & $J$                       & \multirow{2}{*}{$\mathcal{O}\left(N_t L\left(\lVert \bm{\mathcal{C}}_{l}\rVert\right) ^{DN_t}\right)$} &  \\
		SBMA (JMPA)           & $\lVert \mathbf{f}_k \rVert\times N_t$ & $\frac{N_t K_U J}{N_t+K_U-1}$ &                                                                                                        &  \\ \cline{1-4}
		\toprule[1.5pt]
	\end{tabular}
\end{table*}

This section compares SBMA, BIA, and STBC-SCMA in terms of BER and the decoding complexity. We assume that all channel state matrices $\mathbf{h}_k^{j}(m)$ follow the Rayleigh distribution, and all users are at the same distance $d_k=1$ from the base station. In the SBMA scheme, we assume signals for each superuser are allocated to one user. A commonly used codebook is utilized in SCMA and SBMA \cite{CodebookPPT}. {The iteration number of the MPA-based decoder is set to 12. This value is empirical and not necessarily optimal. In traditional	algorithms, a simple approach to determine the iteration number is based on the convergence speed threshold of estimation values during the iterative process \cite{santana2016AReviewOfMPA}. In our future work, we will conduct an extensive investigation into the impact of the iteration number from the perspectives of both BER performance and computational complexity.}



\subsection{BER comparison}

\begin{figure}
	\centering
	\includegraphics[width=3.2in]{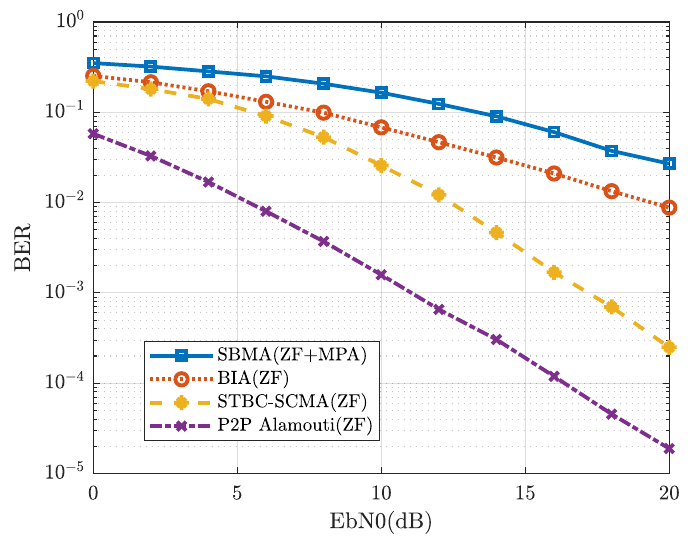}
	\caption{Simulated BER Comparison (Zero-forcing+MPA, $K=L=6, J=4$)}
	\label{fig:BER_Comp_ZF}
\end{figure}

\begin{figure}
	\centering
	\includegraphics[width=3.2in]{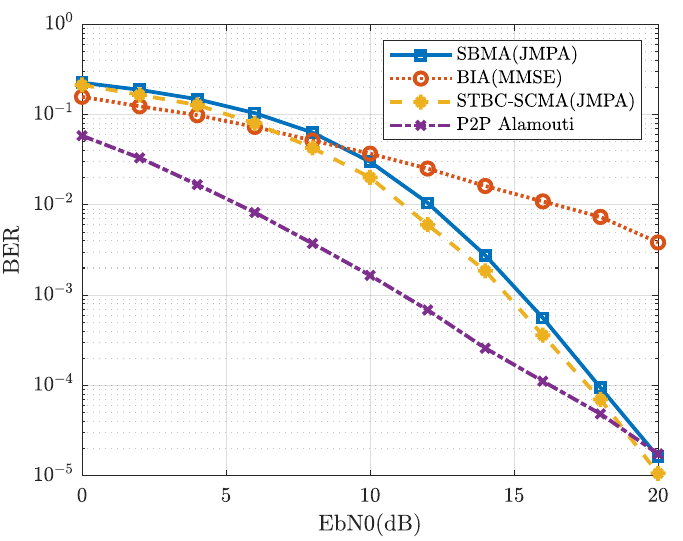}
	\caption{Simulated BER Comparison (JMPA, $K=L=6, J=4$)}
	\label{fig:BER_Comp_JMPA}
\end{figure}

\begin{figure}
	\centering
	\includegraphics[width=3.2in]{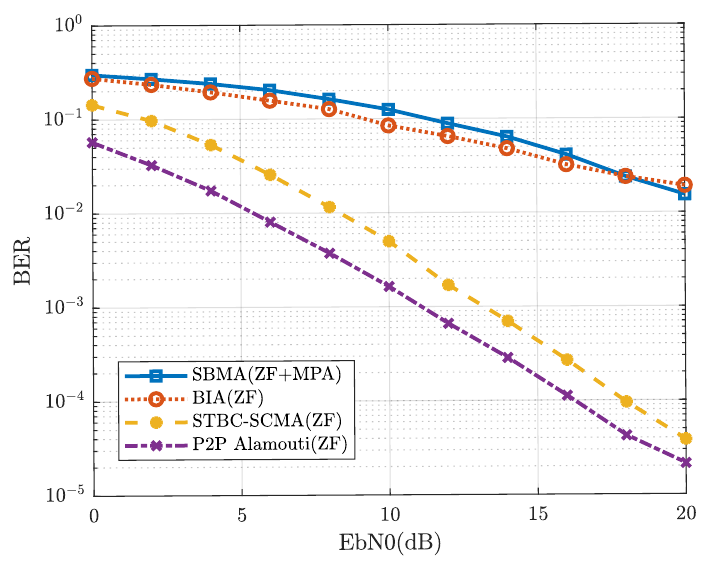}
	\caption{Simulated BER Comparison (Zero-forcing+MPA, $K=L=10, J=5$)}
	\label{fig:BER_Comp_ZF_5T10U}
\end{figure}

\begin{figure}
	\centering
	\includegraphics[width=3.2in]{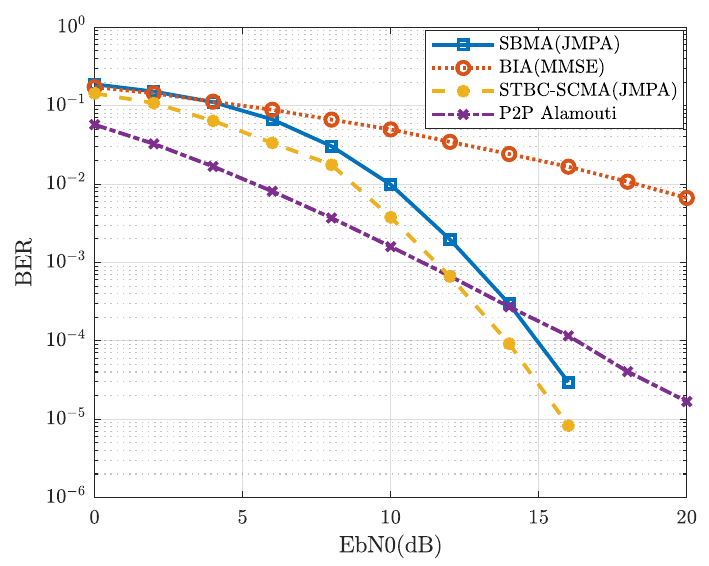}
	\caption{Simulated BER Comparison (JMPA, $K=L=10, J=5$)}
	\label{fig:BER_Comp_JMPA_5T10U}
\end{figure}

\begin{figure}
	\centering
	\includegraphics[width=3.2in]{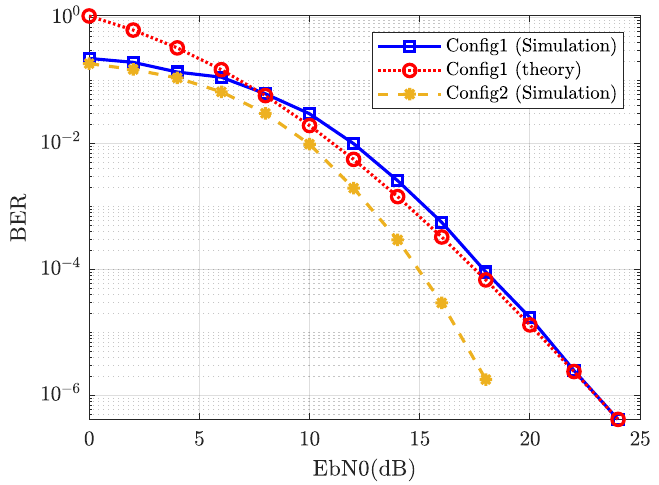}
	\caption{BER comparison (Simulation. vs Theoretical, $K=L=6, J=4$ )}
	\label{fig:theroonlyberjmpabpsk0319024}
\end{figure}

In this section, the error performance of SBMA is compared for BIA and STBC-SCMA. Additionally, we consider an Alamouti scheme in a point-to-point $2\times 1$ MISO system as a benchmark in our simulation. Two system configurations are considered: 1) $K=L=6, J=4$, and 2) $K=L=10, J=5$. Note that in both configurations, we assume $N_t=2$. Moreover, as the number of antennas $N_t$ increases, the coherence time needed for BIA encoding becomes extremely large. While advanced BIA designs can be used to relax the requirement on coherence time, this is beyond the main purpose of this paper. For more discussion about the generalization design for $N_t>2$, please refer to our ongoing work.

Fig. \ref{fig:BER_Comp_ZF} and Fig. \ref{fig:BER_Comp_JMPA} present the BER performance for Configuration 1, while Fig. \ref{fig:BER_Comp_ZF_5T10U} and Fig. \ref{fig:BER_Comp_JMPA_5T10U} present the results for Configuration 2. In Fig. \ref{fig:BER_Comp_ZF} and Fig. \ref{fig:BER_Comp_ZF_5T10U}, both SBMA and STBC-SCMA use the two-stage ZF+MPA decoder, while the BIA and P2P Alamouti schemes utilize the ZF decoder.
In Fig. \ref{fig:BER_Comp_JMPA}  and Fig. \ref{fig:BER_Comp_JMPA_5T10U}, both SBMA and STBC-SCMA use a JMPA decoder, and the BIA scheme employs an MMSE decoder.

In both Fig. \ref{fig:BER_Comp_ZF} and Fig. \ref{fig:BER_Comp_ZF_5T10U}, it is evident that STBC-SCMA significantly outperforms both BIA and SBMA when ZF decoding is considered. This superiority is achieved because STBC with ZF decoding provides spatial diversity, but BIA and SBMA do not. It is important to note that the two-stage decoder for STBC-SCMA and SBMA fails to achieve diversity over tones due to inadequate noise handling in the MPA component, which does not effectively utilize channel state information. This inadequate handling of noise results in poorer performance for SBMA compared to BIA. Furthermore, as $K$ increases in Fig. \ref{fig:BER_Comp_ZF_5T10U}, the length of supersymbols in BIA also increases, leading to higher noise levels in the BIA decoder. Consequently, BIA's performance in Fig. \ref{fig:BER_Comp_ZF_5T10U} is worse than that in Fig. \ref{fig:BER_Comp_ZF}. Additionally, as illustrated in (\ref{EbPower}), the symbol power in MPA decoders depends on the overloading factor $L/J$. Thus, with a larger $L/J$ in Configuration 2, the BER performance of SBMA and STBC-SCMA improves in Fig. \ref{fig:BER_Comp_ZF_5T10U} compared to Fig. \ref{fig:BER_Comp_ZF}.

In Fig. \ref{fig:BER_Comp_JMPA} and Fig. \ref{fig:BER_Comp_JMPA_5T10U}, SBMA demonstrates performance similar to that of STBC-SCMA, which performs better than in Fig. \ref{fig:BER_Comp_ZF_5T10U} and Fig. \ref{fig:BER_Comp_ZF}. These figures illustrate that the JMPA effectively capitalizes on both spatial and tone diversity, since the appropriate utilization of CSI. In our simulation settings, both SBMA and STBC-SCMA achieve a diversity order of 4, with 2 originating from spatial diversity and the remaining 2 from tone diversity. As analyzed in Fig. \ref{fig:BER_Comp_ZF} and Fig. \ref{fig:BER_Comp_ZF_5T10U}, STBC-SCMA and SBMA exhibit improved performance in Fig. \ref{fig:BER_Comp_JMPA_5T10U} compared to Fig. \ref{fig:BER_Comp_JMPA}.

Fig. \ref{fig:theroonlyberjmpabpsk0319024} summarizes the BER performance of SBMA, and validates the derived BER expression in (\ref{eq.theroExp}) for Configuration 1. 

Note that our approach for BER analysis becomes computationally prohibitive as the number of users and/or the size of the codebook increases. The BER expressions for Configuration 2 and other more general cases will be addressed in our future work.

The parameters for the approximation are chosen as $N=2$, $a_1=1/12$, $a_2=1/4$, $c_1=1/2$, and $c_2=2/3$ \cite{Bao2017}. From the figure, we see the closed-form expression proposed in this study aligns well with the simulation results, thereby offering a practical formula for evaluating the performance of the SBMA scheme. Additionally, the simulation results for Configuration 2 demonstrate an improvement of approximately 4 dB when the BER is below $10^{-4}$.

From Fig.~\ref{fig:BER_Comp_ZF} to Fig.~\ref{fig:BER_Comp_JMPA_5T10U}, the two-stage decoder for SBMA demonstrates a diversity gain similar to that of BIA, while the JMPA decoder exhibits a diversity gain comparable to that of STBC-SCMA. Additionally, as discussed in Sect.~\ref{DecodeComplexity}, we know that the decoding complexity of the two-stage decoder is significantly lower than that of the JMPA decoder. Therefore, the two-stage decoder is better suited for scenarios where computing resources at the receivers are limited, and the SNR is relatively high.

\subsection{Performance with imperfect CSI}
\begin{figure}
	\centering
	\includegraphics[width=3.2in]{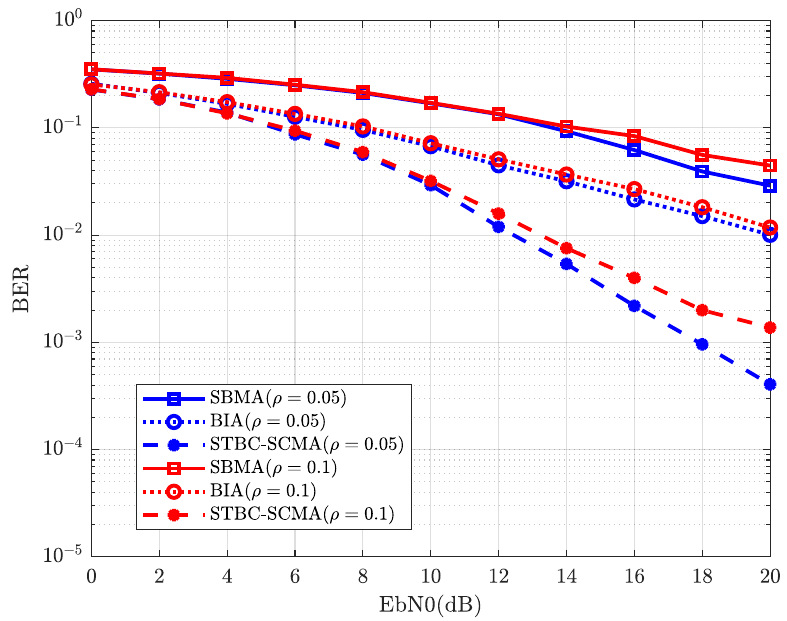}
	\caption{BER comparison (ZF-based decoder with imperfect CSI)}
	\label{fig:ImperfectCSI_ZF}
\end{figure}
\begin{figure}
	\centering
	\includegraphics[width=3.2in]{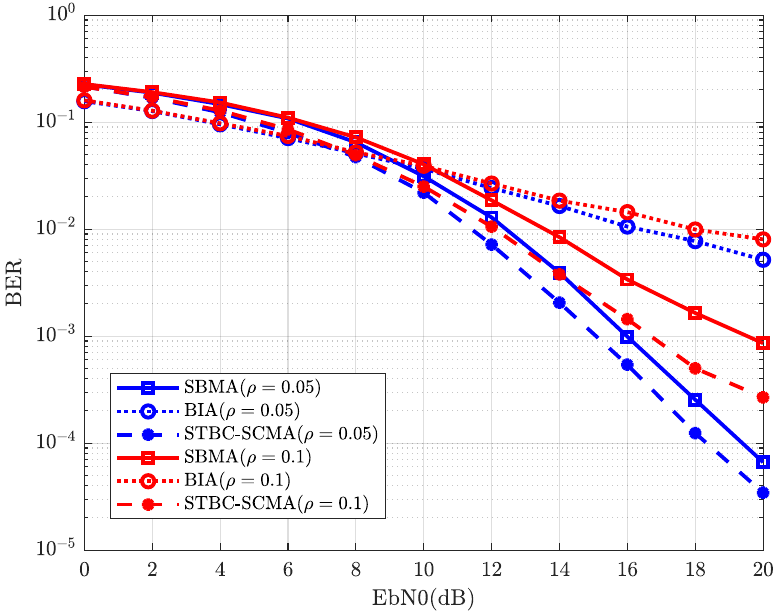}
	\caption{BER comparison (JMPA decoder with imperfect CSI)}
	\label{fig:ImperfectCSI_JMPA}
\end{figure}

This section demonstrates the performance of SBMA, BIA, and STBC-SCMA in the presence of imperfect Channel State Information (CSI). Imperfect CSI arises due to channel estimation errors, which are modeled as \cite{Wang2007OnPerformanceOfMIMOZF}
\begin{equation}
	\bar{\bm{H}} = \bm{H} + \rho\Omega
.
\end{equation}
In this model, $\bm{H}$ represents the real channel matrix, while $\bar{\bm{H}}$ denotes the estimated channel matrix. The estimation error, denoted as $\rho \Omega$, is assumed to be uncorrelated with $\bm{H}$. The entries of $\Omega$ are independent and identically distributed (i.i.d.) zero-mean complex Gaussian random variables with unit variance, and $\rho$ is a parameter that quantifies the accuracy of the channel estimation. As the value of $\rho$ increases, the error in channel estimation also increases.

Fig.~\ref{fig:ImperfectCSI_ZF} and Fig.~\ref{fig:ImperfectCSI_JMPA} illustrate the BER performance for Configuration 1, with $\rho = 0.05$ and $\rho = 0.1$. In Fig.\ref{fig:ImperfectCSI_ZF}, both SBMA and STBC-SCMA utilize the ZF-based two-stage decoder, while BIA employs the ZF decoder. In Fig.\ref{fig:ImperfectCSI_JMPA}, SBMA and STBC-SCMA use the JMPA decoder, whereas BIA employs the MMSE decoder. It is noteworthy that the performance of the schemes in Configuration 2 is similar to that in Configuration 1; therefore, only the results for Configuration 1 are presented in this section to conserve space.

From Fig.~\ref{fig:ImperfectCSI_ZF} and Fig.~\ref{fig:ImperfectCSI_JMPA}, it can be observed that the performance trends of the different schemes under imperfect CSI are consistent with those observed under perfect CSI, as shown in Fig.~\ref{fig:BER_Comp_ZF} and Fig.~\ref{fig:BER_Comp_JMPA}. Furthermore, as $\rho$ increases, the performance of the schemes utilizing JMPA degrades more significantly compared to those employing the ZF-based decoder. This indicates that the JMPA decoder is more sensitive to imperfect CSI, likely due to its operation on a larger factor graph than the ZF-based two-stage decoder. Specifically, in Fig.~\ref{fig:ImperfectCSI_JMPA}, SBMA with the JMPA decoder shows a more considerable performance degradation than STBC-SCMA with the same decoder. This degradation is attributed to noise enhancement after the interference cancellation block during the decoding process, making SBMA with JMPA more sensitive to imperfect CSI. Overall, while SBMA with the two-stage decoder exhibits relatively poorer performance, it demonstrates greater robustness to imperfect CSI. In contrast, SBMA with the JMPA decoder achieves superior performance but is more sensitive to the effects of imperfect CSI.

{
In realistic IoT deployments, the estimation of CSI at receivers is influenced by two main factors: quantization error caused by the limited resolution of analog-to-digital converters (ADCs) and channel estimation error. Previous studies indicate that ADC limitations can induce an additional 1–2 dB loss in BER for traditional schemes \cite{Myers2019MPbasedjointCFOandCE}. To mitigate the impact of CSI imperfections, a larger number of subcarriers $J$  can be utilized, as this provides a higher diversity order, as shown in Fig.~\ref{fig:BER_Comp_JMPA_5T10U}. However, the cost-sensitive nature of IoT devices restricts the feasibility of increasing $J$, which would otherwise elevate computational complexity. Therefore, it is imperative to carefully balance the need for larger $J$ (necessitated by lower CSI accuracy) against the computational complexity constraints imposed by cost considerations. This tradeoff will be systematically investigated in our future work.
}
\subsection{SBMA over other schemes}
\begin{table*}[]
	\centering	
	\caption{Overall comparison on three schemes}
	\label{Tb:ProsAndCons}
	\renewcommand\arraystretch{1.5}	
	\begin{tabular}{cccc}
		\hline
		\toprule[1.5pt]
		& BIA (ML)                      & STBC-SCMA (JMPA) & SBMA (JMPA)                      \\ \hline
		Diversity gain       & Low                       & High      & High                      \\
		Multiplexing gain    & High                      & Low       & High                      \\
		Data streams         & $\frac{N_t J K}{N_t-1+K}$ & $K$        & $\frac{N_t L K_U}{N_t-1+K_U}$ \\
		Coherence Time       & ${(N_t-1)}^K+K({N_t-1})^{K-1}$                      & $N_t$     & ${(N_t-1)}^{K_U}+K_U({N_t-1})^{K_U-1}$                      \\
		Data leakage         & No                        & Yes       & Partial/No                \\
		Unnecessary decoding & No                        & Yes       & Partial/No                \\ \hline
		\toprule[1.5pt]
	\end{tabular}
\end{table*}

Although SBMA originates from BIA and SCMA, it demonstrates superior performance. For comparative analysis, this part examines the enhancements of SBMA over BIA and SCMA, and provides an overall comparison of the three schemes in Table~\ref{Tb:ProsAndCons}.{ SBMA and PD-NOMA are also compared. Detailed explanations are provided as follows.} 
	\begin{enumerate}[]
		\item \textbf{SBMA vs. BIA}. 
		\begin{itemize}
			\item SBMA surpasses BIA in terms of both data stream capacity and user service capability. In Sect.~\ref{modelBIA}, the BIA scheme allows each user to achieve $N_t \times J$ data streams. By integrating the SCMA encoder, SBMA, as detailed in Sect.~\ref{B-SCMA}, supports $N_t \times L$ data streams per superuser, where $L \geq J$. Consequently, SBMA achieves a greater overall data stream capacity. Moreover, SBMA provides enhanced flexibility in distributing data streams among multiple users, accommodating varying user demands efficiently.
            \item SBMA optimizes the length of supersymbols compared to BIA. Because SBMA can support more data streams per slot than BIA, it requires fewer superusers to meet the system's data stream requirements. As a result, shorter supersymbols are achieved.
            \item SBMA surpasses BIA in terms of diversity order by achieving both spatial and tone diversity. Spatial diversity is attained through antenna mode switching during BIA encoding, while tone diversity is achieved through SCMA encoding in SBMA.
		\end{itemize}		
		\item \textbf{SBMA vs. SCMA}. 
		\begin{itemize}
			\item SBMA mitigates data leakage inherent in SCMA-based systems. While SCMA’s MPA decodes both desired and interference signals, risking privacy breaches, SBMA’s BIA encoding and interference cancellation isolate user signals before decoding. This design ensures that only intended signals are processed, thus enhancing privacy.
			\item SBMA eliminates unnecessary decoding. Unlike SCMA, where decoders allocate computing resources to decode unintended data, which is not resource-efficient, SBMA manages unintended data through BIA encoding and the interference cancellation block. As demonstrated in this paper, the IC block effectively eliminates undesired data, enabling SBMA users to filter out unintended signals effortlessly. This feature makes SBMA highly suitable for deployment in downlink multi-user systems.
			\item SBMA achieves a higher multiplexing gain compared to SCMA, as demonstrated in Table~\ref{Tb:OverallComp}. This is facilitated by BIA encoding, which allows SBMA to eliminate interference from unintended superusers' data streams and leverage spatial multiplexing gain for intended transmissions. By contrast, SCMA and STBC do not achieve high multiplexing gain through multiple transmit antennas.
		\end{itemize}		
		\item \textbf{SBMA vs. PD-NOMA}. 
		\begin{itemize}
			\item The proposed SBMA scheme also demonstrates remarkable advantages over conventional PD-NOMA. Unlike PD-NOMA, which requires precise CSIT for effective power allocation, SBMA imposes no mandatory requirement on CSIT to achieve robust interference alignment, making it particularly suitable for resilient transceiver design. Comparative analysis shows that while PD-NOMA suffers significant performance degradation in ultra-dense networks due to multi-user interference accumulation during SIC \cite{Islam2017PDNOMAin5Gsys}, SBMA maintains superior BER performance even under increasing user density. This performance enhancement stems from SBMA’s innovative encoding architecture, which employs BIA to avoid error propagation inherent in PD-NOMA. It should be noted that SBMA does entail a time extension for encoding and noise accumulation introduced by BIA, along with decoding complexity analogous to SCMA. However, such time extension and noise propagation can be substantially mitigated by designing a limited number of superusers, and the computational complexity of decoding can be reduced significantly through advanced methods such as Gaussian-approximated message passing algorithms \cite{Dai2019IterativeGAMP}.
		\end{itemize}
	\end{enumerate}

The table reveals the following: SBMA achieves both high diversity and multiplexing gain, supports more data streams, enhances data privacy, and improves decoding efficiency. Although SBMA requires a longer coherence time, given by ${(N_t-1)}^{K_U} + K_U {(N_t-1)}^{K_U-1}$, compared to STBC-SCMA, it offers a shorter coherence time than BIA due to its reduced number of superusers $K_U$ (with the minimum $K_U$ being 2) compared to normal users. In conclusion, SBMA offers high performance in downlink transmission, making it well-suited for IoT scenarios that demand data privacy, decoding efficiency, and relatively static channel conditions.

\subsection{Typical scenarios analysis}
In this section, we present two typical IoT scenarios to compare the three schemes. The general setup is as follows: A BS equipped with 2 antennas serves $K$ wireless devices, with 4 subcarriers available. For simplicity, we assume the SCMA codebook contains 6 codewords.

\textbf{Scenario 1:} In this scenario, we present an asymptotic analysis of the achievable data streams per slot. Assuming the channel coherence time is sufficiently long, the number of data streams per slot for each scheme can be derived from Table~\ref{Tb:ProsAndCons}. Specifically, BIA achieves $\frac{8K}{1+K}$ data streams per slot, STBC-SCMA achieves 6 data streams (determined by its codebook design), and SBMA achieves $\frac{12K_U}{1+K_U}$ data streams per slot. Given the sufficiently large coherence time, both $K$ and $K_U$ can be scaled to large values. As $K$ and $K_U$ increase, the data streams per slot for BIA and SBMA asymptotically approach 8 and 12, respectively. This analysis demonstrates that SBMA achieves the highest number of data streams under relatively stable channel conditions.

\textbf{Scenario 2:} In this scenario, we compare the schemes with respect to their required coherence time. We assume that both BIA and SBMA achieve at least 6 data streams per slot (equivalent to STBC-SCMA). For BIA, when $K = 3$, it achieves 6 data streams per slot and requires a coherence time of 4 slots. For SBMA, when $K_U = 2$, it achieves 8 data streams per slot and requires a coherence time of 3 slots. In contrast, STBC-SCMA requires a coherence time of 2 slots. This scenario demonstrates that BIA and SBMA necessitate longer coherence times than STBC-SCMA for the same number of data streams, though SBMA requires a shorter coherence time than BIA.  

Based on the analysis of these two scenarios, we conclude that SBMA is well-suited for IoT applications requiring a large number of data streams and relatively stable channel coherence time.

{

As shown in Table~\ref{Tb:ProsAndCons}, the coherence time required for SBMA is given by ${(N_t-1)}^{K_U}+K_U({N_t-1})^{K_U-1}$, which exhibits an exponential dependence on the number of superusers $K_U$. The decoding complexity of SBMA with JMPA follows $\mathcal{O}\left(N_t L\left(\lVert \bm{\mathcal{C}}_{l}\rVert\right) ^{DN_t}\right)$, where the dominant parameters are $L$ and $D$. In massive IoT deployments involving resource-constrained devices, two critical challenges emerge: (1) the coherence time increases exponentially with the number of superusers $K_U$; and (2) systems with a large number of subcarriers ($J > 4$) elevate both $L$ and $D$, thereby increasing the computational burden. To address these scalability issues, we can limit the time extension by establishing an upper bound for $K_U$, regardless of the total number of users $K$. We can then accommodate $K$ users by increasing $J$ and $L$. Subsequently, we can mitigate the computational complexity introduced by $J$ and $L$ by employing advanced message-passing algorithms, such as Gaussian-approximated message-passing algorithms \cite{Dai2019IterativeGAMP}, to align with the computational capabilities of IoT devices. In our future work, we will explore the trade-off between time extension and computational complexity, considering the number of users $K$ and the minimum quality of service (QoS) requirements for each power-limited and low-cost IoT device.
}

\section{Conclusion}
\label{Conc}
This paper proposes SBMA, an effective multiple access method that integrates BIA and SCMA, designed for systems with limited or no CSIT. SBMA enables increased data streams while harnessing the diversity and multiplexing gains inherited from BIA and SCMA. Moreover, SBMA addresses the drawbacks of both BIA and SCMA, such as large supersymbol lengths at BIA and data privacy concerns at SCMA. Two decoders are designed: a two-stage decoder for lower computational complexity and a JMPA decoder for enhanced diversity performance. This paper demonstrates that the flexible framework and decoder options of SBMA accommodate various practical deployment scenarios.

\bibliographystyle{IEEEtran}
\balance

\end{document}